\documentclass[sigconf, nonacm]{acmart}
\AtBeginDocument{%
  }

\usepackage{multirow}
\usepackage{tcolorbox}
\tcbuselibrary{raster,skins,listingsutf8,breakable}
\usepackage{array}
\usepackage{arydshln}
\usepackage{listings}
\usepackage{makecell}

\newtcolorbox{boxposition}[1][]
{enhanced,
attach boxed title to top left=
{xshift=3mm, yshift*=-\tcboxedtitleheight/2},
    colbacktitle=purple!80, 
    colframe=purple!80, 
    colback=purple!5, 
  breakable, 
  title=Position
  }

\newtcolorbox{boxobservation}[1][]{
    enhanced,
    breakable, 
    colbacktitle=blue!60, 
    colframe=blue!60, 
    colback=blue!5, 
    title=Observation
}

\newtcolorbox{boxtakeaway}[1][]{
    enhanced,
    breakable, 
    colbacktitle=violet!80, 
    colframe=violet!80, 
    colback=violet!5, 
    title=Takeaway
    }

\newtcolorbox{boxprompt}[1][]
{enhanced,
  colback=white!3!, 
  breakable, 
  title=Prompt example
  }

\begin{document}

\title{AI Security Map: Holistic Organization of AI Security Technologies and Impacts on Stakeholders}



\author{Hiroya Kato}
\affiliation{%
  \institution{KDDI Research, Inc.}
  \streetaddress{Ohara 2-1-15}
  \city{Fujimino}
  \state{Saitama}
  \country{Japan}
}

\author{Kentaro Kita}
\affiliation{%
  \institution{KDDI Research, Inc.}
  \streetaddress{Ohara 2-1-15}
  \city{Fujimino}
  \state{Saitama}
  \country{Japan}
}

\author{Kento Hasegawa}
\affiliation{%
  \institution{KDDI Research, Inc.}
  \streetaddress{Ohara 2-1-15}
  \city{Fujimino}
  \state{Saitama}
  \country{Japan}
}

\author{Seira Hidano}
\affiliation{%
  \institution{KDDI Research, Inc.}
  \streetaddress{Ohara 2-1-15}
  \city{Fujimino}
  \state{Saitama}
  \country{Japan}
}

\renewcommand{\shortauthors}{H. Kato et. al}

\begin{abstract}
As the social implementation of AI has been steadily progressing, research and development related to AI security has also been increasing.
However, existing studies have been limited to organizing related techniques, attacks, defenses, and risks in terms of specific domains or AI elements.
Thus, it extremely difficult to understand the relationships among them and how negative impacts on stakeholders are brought about. 
In this paper, we argue that the knowledge, technologies, and social impacts related to AI security should be holistically organized to help understand relationships among them.
To this end, we first develop an AI security map that holistically organizes interrelationships among elements related to AI security as well as negative impacts on information systems and stakeholders.
This map consists of the two aspects, namely the information system aspect (ISA) and the external influence aspect (EIA). 
The elements that AI should fulfill within information systems are classified under the ISA.
The EIA includes elements that affect stakeholders as a result of AI being attacked or misused. 
For each element, corresponding negative impacts are identified.
By referring to the AI security map, one can understand the potential negative impacts, along with their causes and countermeasures.
Additionally, our map helps clarify how the negative impacts on AI-based systems relate to those on stakeholders. 
We show some findings newly obtained by referring to our map.
We also provide several recommendations and open problems to guide future AI security communities.
\end{abstract}

\ccsdesc[500]{Security and privacy~Social aspects of security and privacy}
\ccsdesc[300]{Computing methodologies~Artificial intelligence}

\keywords{AI Security, Privacy, Compromise of AI, Misuse of AI, Social impact}


\maketitle

\section{Introduction}
As AI is increasingly utilized in society, research and development on AI security is also further accelerating.
AI security is not only closely connected to the traditional elements of information security, namely confidentiality, integrity, and availability (CIA) but is also strongly related to other AI elements, such as explainability and fairness. 
Accordingly, concerns are growing regarding the complex negative impacts on stakeholders such as individuals and society. 
In other words, the scope of AI security has now extended beyond AI itself to encompass individuals and society, and these are also closely interconnected.
Nevertheless, most existing survey papers \cite{chen2024security, ramirez2022poisoning, hu2021artificial, zhang2024backdoor, mothukuri2021survey} and systematization-of-knowledge (SoK) papers \cite{li2023sok,wingarz2024sok, dibbo2023sok, noppel2024sok} have been limited to the classification of attack and defense techniques for AI.
Some studies systematically investigate techniques or risks in terms of specific AI elements such as explainability \cite{dwivedi2023explainable,mersha2024explainable,schwalbe2024comprehensive}, fairness \cite{kheya2024pursuit,zhang2024ai,parraga2025fairness}, and privacy \cite{lee2024deepfakes,golda2024privacy,chang2024sok,rigaki2023survey}, and examine real-world risks or social impacts \cite{weidinger2022taxonomy,slattery2024ai,pankajakshan2024mapping,aimeur2023fake,castagnaro2024offensive,schroer2025sok}.
However, the relationships among multiple elements including those on AI security have not comprehensively been organized, and they are exclusively within limited domains such as LLMs.
In other words, most existing studies have been limited to organizing related techniques, attacks, defenses, and social impacts from distinct perspectives of fields or AI elements.
Our position is that \textbf{the knowledge, technologies, and social impacts related to AI security should be holistically organized to help understand relationships among them}. 
To the best of our knowledge, there is no study that holistically organize the relationships between AI elements or provide an inclusive classification that covers negative impacts on stakeholders. 

Thus, in this paper, we develop an AI security map that holistically organizes these relationships through examining these interconnections and impacts.
This map consists of the two aspects, namely the information system aspect (ISA) and the external influence aspect (EIA). 
The elements that AI should fulfill within information systems are classified under the ISA. 
On the other hand, the EIA includes elements that affect individuals and society. 
Our map organizes the elements related to AI security, as well as how individuals and society can be affected as a result of AI being attacked or misused.
To develop our map, we surveyed existing papers to classify and organize attacks that can compromise the CIA.
Furthermore, we organized other elements affected by the compromise of the CIA and considered negative impacts that may arise from the ISA.
Similarly, we considered elements in the EIA on the basis of the negative impacts in the ISA and existing papers.
For each element in the EIA, we organized negative impacts caused by the compromise or misuse of AI.
Finally, attacks, causal factors, defenses, and related elements are mapped to each negative impact.
By referring to our map, one can understand the potential negative impacts on AI-based information systems, along with defense and countermeasures necessary to prevent such impacts.
Additionally, our AI security map helps clarify how these negative impacts on the information systems influence individuals and society.
Our contributions are as follows: 
\begin{itemize}
    \item We holistically organize not only multiple elements related to AI, but also those impacting individuals and society.
    \item We consider the mutual relationships between the ISA and the EIA, clarifying how negative impacts on individuals and society are caused by those on information systems.
    \item Our map organizes not only negative impacts caused by attacks on AI but also ones resulting from AI misuse.
    \item In contrast to prior work that has primarily focused on real-world risks or social impacts within specific domains, our work considers them in the broader context of AI security as a whole.
    In particular, we also focus on the subsequent influence resulting from particular risks or impacts.
\end{itemize}

\section{Current Landscape of AI Security}
As AI technologies are incorporated in various fields, concepts of AI security are diversified and become complex more and more.
In such situations, researchers are working on various research directions, such as privacy and explainability, related to AI security.
To present a systematic critical review on huge studies, main approaches, and evaluation methods in these areas, there are many survey papers and SoK papers.
In what follows, we briefly introduce existing survey and SoK papers on these related research fields.

\subsection{Attack and Defense}
In the domain of AI security, there are a huge number of studies regarding attacks and defenses.
Since providing taxonomies or systematic reviews of attacks and defenses is a primary way to realize clear overviews, existing survey papers \cite{chen2024security, ramirez2022poisoning, hu2021artificial, zhang2024backdoor, mothukuri2021survey} and SoK papers \cite{li2023sok,wingarz2024sok, dibbo2023sok, noppel2024sok} mainly focus on organizing attacks on AI or defenses.
We consider that there are three types of taxonomies in existing studies classifying attacks and defenses.

\noindent\textbf{Taxonomy in AI systems.}
The first type classifies attacks and defenses in AI systems.
AI systems mean information systems using AI.
For example, to represent an overview for AI security, \citet{hu2021artificial} review the challenges and research advances for security issues in AI. 
In that work, the lifecycle of an AI system is used as a guide so as to introduce the security threats that emerge at each stage and corresponding countermeasures.
Also, \citet{chen2024security} present a review regarding the security for machine learning (ML) based software systems in light of the fact that no literature review is aimed at a comprehensive investigation of ML-based software systems from the secure development aspect.
\citet{kiribuchi2025securingaisystemsguide} provide an overview of adversarial attacks on AI systems.
They identify 11 major attack types and their resulting impacts.
One of the distinguishing features of that work is focusing on the impacts of attacks on AI systems and providing intuitive visual summaries.

\smallskip
\noindent\textbf{Taxonomy in specific attacks.}
The second type provides a taxonomy with focus on specific attacks.
There are studies that present a comprehensive summary in a specific type of attack on AI, such as poisoning attacks \cite{ramirez2022poisoning}, backdoor attacks \cite{zhang2024backdoor}, and jailbreak attacks \cite{lu2024autojailbreak}.
\citet{ramirez2022poisoning} conduct a survey with highlighting the most relevant information related to poisoning attacks.
That work compiles the most relevant insights and findings found in the latest existing literature regarding poisoning attacks and several defense techniques. 
\citet{zhang2024backdoor} present a comprehensive and systematic summary of both backdoor attacks and defenses targeting multi-domain AI models so as to systematically analyze shortcomings of existing research and address the lack of comprehensive reviews.
\citet{lu2024autojailbreak}  propose AutoJailbreak, a framework designed to comprehensively evaluate the resilience of LLMs against jailbreak attacks.
Furthermore, that work conducts a comprehensive examination of jailbreak attacks and defenses.
In total, over 28 jailbreak attacks and 12 jailbreak defenses are organized.

\smallskip
\noindent\textbf{Taxonomy in specific domains.}
The third type deals with the classification of attacks and defenses in specific domains.
There are studies \cite{mothukuri2021survey,shen2022sok,wingarz2024sok} that summarize related work in order to provide taxonomy in specific domains.
\citet{shen2022sok} perform the first systematization of knowledge of growing AI security research in the field of autonomous driving (AD). 
They collect and analyze 53 existing papers, and systematically taxonomize them on the basis of research that is critical for the security field.
That work organizes knowledge from the perspective of important research aspects, such as AI components related to the attack and defense as well as evaluation methodologies.
To address the existing gaps in understanding the complexities of Edge AI, \citet{wingarz2024sok} provide a comprehensive survey of the challenges necessary for enhancing the security and safety of Edge AI.
They examine both existing threats and their relevant countermeasures with social implications. 
Finally, they identify a series of open research challenges and present required action to advance solutions in the area.
\begin{boxobservation}
Most existing studies just provide taxonomies of attacks and defenses from the perspective of individual research.
Thus, there has been no description of what specific negative impacts in the real world can be caused by each attack.
\end{boxobservation}

\subsection{Elements related to AI Security}
In recent years, as AI is expected to be applied in various fields, AI related elements, such as fairness and privacy, have become increasingly important.
It is crucial to consider whether AI meets these elements closely related to AI security because they can be compromised by attacks on AI, which may have significant negative impacts.
Existing studies have summarized the various risks, challenges, and desired research directions in terms of each element entailed by the utilization of AI. 
In what follows, we introduce related studies that address representative elements in AI.

\smallskip
\noindent\textbf{Explainability.}
As AI increasingly exerts an influence on human decision-making, it is becoming essential to understand the rationale behind AI predictions. 
To this end, Explainable AI (XAI) has been studied.
XAI refers to technologies that enable the explanations of the reasoning and processes by which AI models make decisions or predictions in a manner comprehensible to humans. 
The advancement of XAI is expected to enhance the trustworthiness and transparency of AI predictions.
There are various systematic studies \cite{dwivedi2023explainable,mersha2024explainable,schwalbe2024comprehensive} on the field of XAI.
For example, \citet{dwivedi2023explainable} survey programming techniques for XAI and present the different phases of XAI in a typical ML development process. 
The various XAI approaches are classified to discuss the key differences among the existing XAI techniques.
\citet{schwalbe2024comprehensive} provide a complete taxonomy of XAI methods with respect to notions included in the existing research.

\smallskip
\noindent\textbf{Fairness.}
To prevent AI models from perpetuating the biases present in the data and producing unfair decisions, researchers has worked on the research on fairness in AI. 
Although fairness is a relatively new research area, as of now, there are multiple studies \cite{kheya2024pursuit,zhang2024ai,parraga2025fairness} in this area.
For example, \citet{zhang2024ai} review recent advances in AI fairness aimed at bridging gaps for practical deployment in real-world scenarios. 
That review seeks not only to identify existing gaps but also to propose solutions that reconcile the theoretical underpinnings of fairness with the complex realities of real-world data dynamics.
Moreover, they highlight the limitations and significant potential for real applications.
\citet{parraga2025fairness} provide an in-depth overview of representative debiasing methods for fairness-aware neural networks in vision and language domains. 

\smallskip
\noindent\textbf{Privacy.}
Privacy is one of the most related domains to AI security because it is a key principle for developing ethical and secure AI.
In studies \cite{lee2024deepfakes, golda2024privacy}, AI privacy risks are presented.
\citet{golda2024privacy} conduct a meticulous examination of the privacy and security challenges inherent to Generative AI.
That study provides five pivotal perspectives essential for a comprehensive understanding of these intricacies. 
\citet{lee2024deepfakes} present 12 high-level privacy risks that AI technologies either newly created or exacerbated.
There are also papers that systematically organize privacy risks or provide taxonomy of attacks.
\citet{chang2024sok} provide a systematic overview of attacks on healthcare AI to facilitate privacy leakage and defenses in response to inconsistent settings in terms of healthcare deployment scenarios and threat models.
\citet{rigaki2023survey} propose an privacy attack taxonomy, together with a threat model that allows the categorization of different attacks based on the adversarial knowledge.
They also offer an overview of defenses and discussion of the open problems and future directions.
\begin{boxobservation}
Existing studies systematically investigate techniques or risks in terms of specific AI elements. 
However, the relationships among multiple elements including those on AI security have not systematically been organized.
\end{boxobservation}

\subsection{Real-World Risks and Impacts}
As another direction, some studies also discuss what risks and impacts \cite{weidinger2022taxonomy,slattery2024ai,pankajakshan2024mapping} are brought about by the compromise or misuse of AI \cite{schroer2025sok,aimeur2023fake,castagnaro2024offensive} in the real world.

\smallskip
\noindent\textbf{Social impact of AI.}
\citet{weidinger2022taxonomy} develop a comprehensive taxonomy of ethical and social risks associated with language models (LMs). 
They identify 21 risks appearing in current LMs and develop a taxonomy consisting of 6 risk areas to help understand their landscape.
That work shares foresight to help make the landscape of risks associated with LMs easier to parse, which contributes to guiding required action to address these risks.
\citet{slattery2024ai} present several implications for the collective understanding of how the landscape of AI risks is constructed.
That study highlights the need for a balanced approach that both drives technological progress and embraces social responsibility by considering the impact of AI on the workforce, economic dynamics, and ethical issues.
\citet{pankajakshan2024mapping} discuss LLM security and stakeholder risks.
They argue that organizations are deploying LLM-integrated systems without understanding the severity of potential consequences. 
Moreover, whereas existing studies by OWASP and MITRE offer a general overview of threats and vulnerabilities, they pointed out that there is a few methods for directly and succinctly analyzing the risks for security practitioners, developers, and key decision-makers who are working with this novel technology. 
To address the limitations, they finally propose a risk assessment process.

\smallskip
\noindent\textbf{Misuse of AI.}
AI can have adverse impacts on people and society regardless of attacks on AI.
In addition to the attack on AI, the misuse of AI is one of the factors that can have negative impacts on the real world.
This concern is discussed in several studies \cite{schroer2025sok,aimeur2023fake,castagnaro2024offensive}.
\citet{aimeur2023fake} mention that fake news can have a significant impact on society, and false content is easier to generate and harder to detect by using AI based tools.
Recent studies \cite{castagnaro2024offensive,schroer2025sok} discuss a emerging paradigm that integrates AI technologies to conduct or enhance cyber attacks, which is called offensive AI (OAI).
\citet{castagnaro2024offensive} explore whether AI can enhance the directory enumeration process and propose a novel LM-based framework. 
On the other hand, \citet{schroer2025sok} devise a common set of criteria reflecting essential technological factors related to OAI.
They consider OAI as the crucial means to violate security and privacy.
In their studies, humans and society are considered and summarized as targets of OAI attacks.
\begin{boxobservation}
Some studies discuss impacts on stakeholders beyond merely classifying attacks and defenses. 
However, they discuss real-world risks or social impacts exclusively within limited domains such as LLMs and OAI. 
\end{boxobservation}

\subsection{Limitation on Current Landscape}
Most systematic studies on AI security have primarily organized knowledge from a technical perspective by classifying attacks on AI and defense methods.
On the other hand, techniques or some risks related to AI elements are also discussed in existing studies.
However, they focus on individual elements, such as fairness and privacy.
Thus, the relationships between the elements and AI security have not been discussed or organized despite their close and important relationship.
Some papers discuss impacts on stakeholders or society in addition to the classification of attacks and defenses.
However, their overviews have been limited to real-world risks or impacts only in specific fields. 
Furthermore, the subsequent influence resulting from particular risks or impacts have not been considered.
It is important to consider such chains of impacts so as to understand potential impacts on people and society in practice.
Today, AI has a diverse impacts not only on information systems, but also on stakeholders.
Consequently, the scope of AI security has expanded beyond a limited group of researchers to encompass most people including engineers, users, and even non-users. 
In a nutshell, it is no longer sufficient to examine individual elements in isolation, which requires to take the broader context into account.

\section{Holistic Organization of AI Security}
\subsection{Position Statement}
As described in the previous section, most existing studies have been limited to organizing related techniques, attacks, defenses, and social impacts from distinct perspectives or AI elements.
Hence, it is extremely difficult to understand the relationships among related elements by referring to multiple papers in their current state. 
As AI becomes increasingly integrated into information systems and society, a holistic overview is required not only from the standpoints of research but also from those of business and general public in order to comprehensively understand potential impacts caused by the compromise or misuse of AI.
In particular, in this increasingly complex field, interrelationships among all elements associated with AI security should be comprehensively organized.
Furthermore, it is essential to clarify how negative impacts on individuals and society are brought about.
In short, our position is summarized as follows:
\begin{boxposition}
The knowledge, technologies, and social impacts related to AI security should be holistically organized to help understand relationships among them.
To this end, it is necessary not only to classify attacks and defenses, but also to organize the interrelationships among related elements, as well as how individuals and society may be influenced by attacks or the misuse of AI.
\end{boxposition}
In contrast to the previous studies, our work considers and summarizes the interrelationships between those elements through negative impacts, in addition to the classification of various types of attacks and defenses.
In what follows, we propose new holistic organization that encompasses AI security technologies and negative impacts on information systems and stakeholders.

\subsection{AI Security Map}
\begin{table*}[t]
\centering
\caption{AI security map for negative impacts related to CIA in the ISA.}
\label{tab:ISA_CIA}
\LARGE
\scalebox{0.52}{
    \begin{tabular}{|c|p{0.25\textwidth}|p{0.3\textwidth}|p{0.38\textwidth}|p{0.27\textwidth}p{0.4\textwidth}|} \hline

    \multirow{1}{*}{\textbf{Elements}} & \multirow{1}{*}{\textbf{Negative impacts}} & \multirow{1}{*}{\textbf{Attack or cause}} & \multirow{1}{*}{\textbf{Defenses or countermeasures}} & \multicolumn{2}{c|}{\textbf{Related elements in the EIA}} \\ \hline

    \multirow{15}{*}{Confidentiality} & Training data leakage & - Membership inference attack \cite{shokri2017membership} & - Differential privacy (DP) \cite{abadi2016deep}\par - Encryption technology \cite{juvekar2018gazelle}\par - AI access control & - Privacy\par  - Copyright and authorship\par  - Reputation & - Psychological impact\par - Compliance with laws and regulations \\ \cline{2-6}

    & Personal information leakage & - Membership inference attack \cite{shokri2017membership}\par - Prompt injection \cite{shen2024anything} & - DP \cite{abadi2016deep}\par - Federated learning \cite{wang2020federated}\par - Personal information masking\par - AI access control& - Privacy\par - Copyright and authorship \par - Reputation & - Psychological impact \par - Compliance with laws and regulations \\ \cline{2-6}

    & Reconstruction of training data & - Model inversion attack \cite{zhang2020secret} & - DP \cite{abadi2016deep}\par - Encryption technology \cite{juvekar2018gazelle}\par - AI access control & - Privacy\par - Copyright and authorship\par - Reputation & - Psychological impact\par - Compliance with laws and regulations \\ \cline{2-6}

    & Model information leakage & - Model extraction attack \cite{tramer2016stealing} & - DP \cite{abadi2016deep}\par - Detection of model extraction attack\par - AI access control & - Privacy\par - Copyright and authorship\par  - Reputation & - Psychological impact\par - Compliance with laws and regulations \\ \cline{2-6}

    & Leakage of system prompts & - Prompt leaking \cite{hui2024pleak}& - Prompt checking & - Reputation & \\ 
    \hline

    \multirow{17}{*}{Integrity} & Manipulation of AI output for specific inputs & - Adversarial examples \cite{szegedy2013intriguing,papernot2016limitations} & - Adversarial training \cite{madry2017towards}\par - Detection of adversarial examples \cite{tramer2022detecting} \par - Certified robustness (CR) \cite{li2023sok} & - Reputation \par - Disinformation\par - Usability\par - Consumer fairness & - Compliance with laws and regulations\par - Critical infrastructure\par - Physical impact\par - Medical care \\ \cline{2-6}
    
    & Degradation of AI performance due to training data contamination & - Poisoning attack \cite{carlini2024poisoning,biggio2012poisoning}& - Detection of poisoned data\par - CR \cite{li2023sok} & - Reputation \par - Misinformation\par - Usability\par - Consumer fairness & - Compliance with laws and regulations\par - Critical infrastructure\par - Physical impact\par - Medical care \\ \cline{2-6}

    & Manipulation of AI output under specific conditions & - Backdoor attack \cite{gu2017badnets,saha2020hidden} & - Detection of triggers \cite{doan2020februus}\par - Detection of backdoored models \cite{xu2021detecting}\par - CR \cite{li2023sok}& - Privacy\par - Disinformation\par - Usability\par - Consumer fairness\par - Reputation\par - Human-centric principle & - Compliance with laws and regulations\par  - Physical impact\par - Ethics\par - Economy \par - Critical infrastructure\par - Medical care \\ \cline{2-6}

    & Generation of harmful responses & - Prompt injection \cite{shen2024anything} & - Toxicity detection \cite{hu2024toxicity} \par - Prompt checking & - Privacy\par - Disinformation\par - Consumer fairness & - Reputation \par - Compliance with laws and regulations\par - Ethics \\ 
    \hline

    \multirow{18}{*}{Availability} & Misclassification by AI, leading to degradation of functionality or service quality & - Adversarial examples \cite{szegedy2013intriguing,papernot2016limitations} & - Adversarial training \cite{madry2017towards}\par - Detection of adversarial examples \cite{tramer2022detecting} \par - CR \cite{li2023sok}& - Reputation \par - Human-centric principle\par - Ethics \par - Critical infrastructure & - Compliance with laws and regulations\par - Physical impact\par  - Economy \par - Medical care \\ \cline{2-6}

    & Continuous decrease in predictive accuracy, leading to degradation or cessation of functionality or service quality & - Poisoning attack \cite{carlini2024poisoning,biggio2012poisoning}& - Detection of poisoned data\par - CR \cite{li2023sok}& - Reputation\par - Usability\par - Physical impact \par - Psychological impact&  - Financial impact\par - Economy\par - Critical infrastructure\par - Medical care \\ \cline{2-6}

    & AI output manipulated under specific conditions, leading to degradation of functionality or service quality & - Backdoor attack \cite{gu2017badnets,saha2020hidden} & - Detection of triggers \cite{doan2020februus}\par - Detection of backdoored models \cite{xu2021detecting}\par - CR \cite{li2023sok}& - Reputation\par - Usability\par - Physical impact \par - Psychological impact&  - Financial impact\par - Economy\par - Critical infrastructure\par - Medical care \\ \cline{2-6}

    & Service disruption due to high consumption of AI resources & - Model DoS \cite{zhang2024safeguard}& - Token limit\par - AI access control & - Reputation\par - Physical impact\par - Psychological impact \par - Financial impact & - Economy\par - Critical infrastructure\par - Medical care \\ 
    \hline
    \end{tabular}
}
\end{table*}
We develop new holistic organization of AI security technology and impacts on information systems and stakeholders, which is called AI security map.
This map consists of the two aspects, namely ISA and EIA.
The elements that AI should fulfill within information systems are classified under the ISA.
On the other hand, the EIA includes elements that affect individuals and society as a result of AI being attacked or misused. 
The definitions of the elements in the ISA and the EIA are shown in Table~\ref{tab:info_def} and Table~\ref{tab:ext_def}, respectively, in Appendix~\ref{apdx:def_aspects}.
For each element, corresponding negative impacts are identified.
The negative impacts in the ISA are mainly attributed to attacks on AI. 
They strongly related to information systems using AI.
For example, confidentiality is breached when information systems that a LLM is integrated with are attacked via the prompt injection in order to elicit personal information from LLMs.
On the other hand, the negative impacts categorized under the EIA may arise not only when AI is attacked but also when AI that is functioning properly is misused by malicious users.
This map organizes these negative impacts, the attacks and factors that cause them, as well as the corresponding defense methods and countermeasures.
In addition, we holistically examine the relationships between the ISA and the EIA, clarifying how the compromise or misuse of AI can bring about negative impacts on stakeholders such as individuals and society. 
We assume four types of security targets at this stage.
These security targets mean primary stakeholders that could be affected by AI attacks or misuse.
The definitions of these security targets are shown in Table~\ref{tab:security_target} in Appendix~\ref{apdx:def_target}.

\smallskip
\noindent\textbf{Information system aspect.}
In this aspect, the elements that AI must satisfy within information systems are classified. 
The primary focus is on the three elements of information security, known as CIA.
Other elements are mainly organized based on their being affected by the compromise of CIA. 
By examining the negative impacts on these elements, one can better understand their underlying causes and potential countermeasures. 
Additionally, it is assumed that the negative impacts on the elements classified under this category may influence elements in the EIA.

\smallskip
\noindent\textbf{External influence aspect.}
In this aspect, elements that impact individuals and society, such as privacy violations and infringements of rights like copyright, are classified. 
The negative impacts in this aspect are assumed to arise not only from attacks on AI within information systems but also from the misuse of AI. 
By examining the negative impacts on the elements within this aspect, one can understand which aspects of the information system are compromised and how such damage can affect people and society, as well as potential countermeasures.


\noindent\textbf{Relationship of elements in AI security map.}
The AI security map organizes the relationships between the elements of the ISA and the EIA. 
Specifically, we consider that compromise or misuse of any element within the ISA is related to elements of the EIA. 
Elements within the ISA other than the CIA are affected by breaches of the CIA.
Elements of the EIA are mainly impacted by compromise of elements within the ISA. 
By referring to the AI security map, one can holistically understand these relationships.

\begin{table*}[t]
	\centering
	\caption{AI security map for negative impacts related to elements other than CIA in the ISA.}
	\label{tab:ISA_others}
        \LARGE
	\scalebox{0.50}{
        \begin{tabular}{|c|p{0.2\textwidth}|p{0.15\textwidth}|p{0.25\textwidth}|p{0.35\textwidth}|p{0.28\textwidth}p{0.4\textwidth}|} \hline

         \multirow{2}{*}{\textbf{Elements}} & \multirow{2}{*}{\textbf{Negative impacts}} & \multirow{1}{*}{\textbf{Compromised}} & \multirow{2}{*}{\textbf{Attack or cause}} & \multirow{2}{*}{\textbf{Defenses or countermeasures}} & \multicolumn{2}{c|}{\multirow{2}{*}{\textbf{Related elements in the EIA}}} \\ 
         & & \textbf{elements} & & & & \\ \hline

        \multirow{3}{*}{Explainability} & Difficulty in understanding AI inference results & - Integrity & - Attacks on explainability & - XAI \cite{schwalbe2024comprehensive}\par - Robust explainability & - Reputation\par - Transparency\par - Psychological impact & - Financial impact\par - Economy\par - Medical care \\ 
        \hline

        \multirow{4}{*}{Output Fairness} & Bias in the AI output & - Integrity &  - Bias in training data \cite{shankar2017no} & - Defensive method for integrity\par - Detection of bias in AI output\par -  Elimination of bias in training data\par - Creation of fair AI models & - Usability\par - Consumer fairness\par - Reputation\par - Medical care & - Compliance with laws and regulations\par - Psychological impact\par - Ethics \\ 
        \hline

        \multirow{5}{*}{Safety} & Harm to humans due to decreased prediction accuracy or unexpected behavior by AI & - Integrity & & - Defensive method for integrity\par - Fail-safe mechanism & - Reputation\par - Human-centric principle\par - Physical impact\par - Psychological impact \par - Critical infrastructure &- Compliance with laws and regulations\par  - Financial impact\par - Ethics\par - Economy\par - Medical care \\ 
        \hline
        
        \multirow{6}{*}{Accuracy} & Decrease in AI prediction accuracy & - Integrity & & - Defensive method for integrity & - Misinformation\par - Usability\par - Consumer fairness\par - Reputation\par - Human-centric principle\par - Physical impact & - Psychological impact\par - Financial impact\par - Economy\par - Critical infrastructure\par - Medical care \\ 
        \hline

        \multirow{6}{*}{Controllability} & Unintended output or behavior by administrators & - Integrity & - Adversarial examples \cite{szegedy2013intriguing,papernot2016limitations} \par- Prompt injection \cite{shen2024anything}\par - Indirect prompt injection \cite{greshake2023not} \par - Backdoor attack \cite{gu2017badnets,saha2020hidden}\par - Cyber attack & - Defensive method for integrity & - Disinformation\par - Usability\par - Consumer fairness\par - Reputation\par - Human-centric principle\par - Critical infrastructure & - Compliance with laws and regulations\par - Physical impact\par - Psychological impact\par - Financial impact\par - Economy\par - Medical care \\
        \hline

        \multirow{4}{*}{Trustworthiness} & Difficulty in determining the trustworthiness of AI output & - Integrity \par - Explainability & - Hallucination & - Quantification of uncertainty \cite{ye2024benchmarking} \par - RAG \cite{asai2023self} \par - XAI  \cite{schwalbe2024comprehensive}\par - Detection of hallucination \cite{valentin2024cost} & \par - Usability\par - Reputation\par - Psychological impact\par - Transparency & - Economy\par - Critical infrastructure\par - Medical care \\ 
        \hline

        \end{tabular}

    }
    
\end{table*}
\section{Discussion}
In this section, we discuss new findings by referring to the AI security map.
To be specific, we conduct discussion in order to answer the following questions.
\begin{enumerate}
    \item How are elements or negative impacts in the ISA interrelated?
    \item How are elements or negative impacts in the EIA interrelated?
    \item What impact is produced on each security target?
    \item How does negative impacts affect individuals and society?
    
    
\end{enumerate}
At this stage, we represent the AI security map in the tabular form.
Due to space limitations, the AI security map in terms of the ISA is divided into two parts: one for CIA-related elements and the other for non-CIA elements.
Additionally, as for the EIA, we divide the map into three ones on the basis of security targets.
In what follows, we show the concrete maps to provide some insights obtained by referring to them.

\subsection{Insights into Information System Aspect}
Table~\ref{tab:ISA_CIA} shows the AI security map for negative impacts related to CIA in the ISA.
We identified that the compromise of CIA elements causes 13 negative impacts on information systems.
These negative impacts can directly hinder the functions and operations of AI based information systems. 
Furthermore, we discovered that there are more negative impacts related to confidentiality compared to those on integrity and availability.
This is consistent with the fact that research regarding privacy is actively conducted.
In many cases, the elements of CIA are compromised first by attacks on AI. 
As a result, this can then affect other elements in the ISA.

Table~\ref{tab:ISA_others} shows the AI security map for negative impacts related to elements other than CIA in the ISA.
As to elements other than CIA, we have organized the six elements, namely explainability, output fairness, safety, accuracy, controllability, and trustworthiness.
Among them, some elements can be caused by specific attacks.
However, they can also occur as a result of the compromise of CIA elements.
We identified one negative impact on each of these six elements.
In particular, it is clear that the compromise of integrity can lead to negative impacts on all the six elements.
For example, if integrity is attacked and the expected prediction results are no longer returned, this can affect explainability, making it impossible to provide explanations for the AI’s prediction results. 
From the information security perspective, it is natural that integrity is related to various elements of information systems.
As to trustworthiness, explainability can also bring about the negative impact.
Overall, in the context of AI security, our map made us realize that integrity is a crucial element that affects many other aspects in information systems with AI.

\begin{boxtakeaway}
Confidentiality tends to be the primary targets of attacks in AI based information systems.
Also, integrity is the most influential elements in the ISA.
Keeping integrity intact is extremely difficult but essential, preventing other elements from being compromised.
    
\end{boxtakeaway}

\begin{table*}[t]
\centering
\caption{AI Security Map for negative impacts on consumers in the EIA.}
\label{tab:EIA_consumer}
\huge
\scalebox{0.42}{
    \begin{tabular}{|c|p{0.6\textwidth}|p{0.32\textwidth}|p{0.35\textwidth}|p{0.55\textwidth}|} \hline

    \multirow{2}{*}{\textbf{Elements}} & \multirow{2}{*}{\textbf{Negative impacts}} & \multirow{1}{*}{\textbf{Compromised elements}} &  \textbf{Causal factors or } & \multirow{2}{*}{\textbf{Defenses or countermeasures}} \\ 
    &  & \textbf{in the ISA} & \textbf{related elements in the EIA} &  \\ \hline

    \multirow{5}{*}{Privacy} & Consumers accidentally inputting their personal information into generative AI or similar systems &  - Transparency & - Social engineering attack \cite{schmitt2024digital} & - Anonymization technology \cite{slijepvcevic2021k} \par - DP \cite{abadi2016deep}\par - Federated learning \cite{wang2020federated}\par - Machine unlearning \cite{warnecke2021machine}\par - Encryption technology \cite{juvekar2018gazelle}\\ 
    \hline

    \multirow{5}{*}{Misinformation} & Outputting misinformation by AI &  - Integrity\par  
 - Accuracy\par - Controllability\par - Explainability\par - Trustworthiness & - Poisoning attack on RAG \cite{nazary2025poison} \par - Hallucination & - Defensive method for integrity\par - Data curation \cite{northcutt2021pervasive}\par - RAG \cite{asai2023self}\par - XAI \cite{schwalbe2024comprehensive}\par - Detection of hallucination \cite{valentin2024cost}\\ 
    \hline

    \multirow{5}{*}{Usability} & The decline in the usability of AI &  - Integrity\par - Availability\par - Accuracy\par - Controllability\par - Output fairness & & - Defensive methods for integrity and availability\par - RAG \cite{asai2023self}\\ 
    \hline

    \multirow{8}{*}{Consumer fairness} & Loss of job or life opportunities due to biased AI output &  - Integrity\par - Controllability\par - Output fairness & & - Defensive method for integrity\par - Human in the loop\par - Countermeasures for output fairness\par - Detection of bias in AI output \\  \cline{2-5}

     & Unfair biased and discriminatory output &  - Integrity\par - Controllability\par - Output fairness & & - Defensive methods for integrity\par - AI alignment \cite{ouyang2022training}\par - Countermeasures for output fairness\par - Detection of bias in AI output \\ 
    \hline

    \multirow{4}{*}{Transparency} & Unintentionally using AI & - Explainability & & - Identification of AI-generated output\par - Watermarking for generative AI \cite{kirchenbauer2023watermark} \\ \cline{2-5}

    & Using AI without recognizing the risks & - Explainability\par - Trustworthiness & & - AI-generated output with disclaimers\par - Education and follow-up \\ 
    \hline

    \multirow{4}{*}{Human-centric principle} & Improperly manipulating the decision-making of consumers by AI &  - Integrity\par - Explainability\par - Controllability\par - Output fairness & & - Defensive methods for integrity\par - Human in the loop \\ 
    \hline

    \multirow{2}{*}{Ethics} & Unethical output or actions by AI & - Integrity & - Jailbreak & - Education and follow-up\par - AI alignment \cite{ouyang2022training}\\ 
    \hline

    \multirow{4}{*}{Physical impact} & Physical harm to consumers caused by AI & - Integrity\par - Accuracy\par - Controllability\par - Safety & - Human-centric principle & - Defensive methods for integrity \\ 
    \hline

    \multirow{5}{*}{Psychological impact} & Psychological harm to consumers caused by AI &  - Integrity\par - Availability\par - Controllability\par - Safety\par - Output fairness & - Consumer fairness & - Defensive methods for integrity\par - Defensive methods for availability \\ 
    \hline

    \multirow{6}{*}{Financial impact}& Financial harm to consumers caused by AI & - Integrity\par - Availability\par - Accuracy\par - Controllability\par - Safety \par - Output fairness & - Human-centric principle & - Defensive methods for integrity \par - Defensive methods for availability \\ 
    \hline

    \end{tabular}

}

\end{table*}

\subsection{Insights into External Influence Aspect}
In total, we identified 20 elements and 36 negative impacts in the EIA. 
This indicates that AI security is closely related to a variety of elements and can have a significant impact on individuals and society.
Most negative impacts on the EIA may not only derive from negative impacts in the ISA, but may also be related to other impacts within the EIA.
For example, negative impacts from cyber attacks or disinformation can be the cause of negative economic impacts on non-consumers. 
A common feature among these elements is that the negative impacts caused by the misuse of AI can occur even when the elements of the ISA are satisfied.
These negative impacts do not necessarily affect individuals or society immediately when AI is misused.
Ultimately, when the malicious user achieves their specific objective through the AI misuse, negative impacts are inflicted on individuals and society.
In the following subsections, we provide detailed interpretation of negative impacts on each security target in the EIA.

\begin{boxtakeaway}
There are many negative impacts on individuals and society.
The elements related to the misuse of AI do not necessarily have an immediate impact at the moment they are misused. 
The impact tends to spread when the objective of the misuse is achieved. 
\end{boxtakeaway}

\subsection{Negative Impacts on Each Security Target}
\noindent\textbf{Consumers.}
Table~\ref{tab:EIA_consumer} shows the AI security map for negative impacts on consumers in the EIA.
We identified 12 negative impacts on consumers and 10 related elements. 
It was found that most of the negative impacts on consumers are the result of the compromise of elements in the ISA. 
It is intuitive that negative impacts on consumers, which is users using AI are associated with negative impacts on information systems or AI itself. 
On the other hand, some impacts can be caused directly by specific attacks. 
For example, social engineering attacks \cite{schmitt2024digital} and poisoning attacks on RAG \cite{nazary2025poison} can lead to negative impacts related to privacy and misinformation, respectively. 
Physical, psychological, and financial impacts may occur when human-centered principles are compromised.

\smallskip
\noindent\textbf{Non-consumers.}
Table~\ref{tab:EIA_non-consumer} shows the AI security map for negative impacts on non-consumers in the EIA.
For non-consumers, we identified 13 negative impacts and 10 related elements.
Similar to consumers, many of these negative impacts result from the compromise of ISA elements. 
However, it is important to note that even when ISA elements are functioning properly, there can still be negative impacts if these elements are misused. 
For example, abusing the accuracy or availability of AI can facilitate the spread of misinformation or enable cyber attacks. 
In addition, many negative impacts are caused by attacks exploiting AI or by the compromise of other EIA elements. 
In our map, four negative impacts related to privacy were identified, indicating that there is a high risk of privacy violations for users who do not use AI themselves. 
This demonstrates the significant external influence that AI can have.

\begin{table*}[t]
\centering
\caption{AI Security Map for negative impacts on non-consumers in the EIA.}
\label{tab:EIA_non-consumer}

\huge
\scalebox{0.42}{

    \begin{tabular}{|c|p{0.5\textwidth}|p{0.3\textwidth}|p{0.2\textwidth}|p{0.35\textwidth}|p{0.6\textwidth}|} \hline

    \multirow{2}{*}{\textbf{Elements}} & \multirow{2}{*}{\textbf{Negative impacts}} & \multicolumn{2}{c|}{\textbf{Related elements in the ISA}} &  \textbf{Causal factors or } & \multirow{2}{*}{\textbf{Defenses or countermeasures}} \\ \cline{3-4}
     &  & \textbf{Compromise} & \textbf{Abuse} & \textbf{related elements in the EIA} &  \\ \hline

    \multirow{3}{*}{Cyber attack} & Using AI for cyber attacks & - Confidentiality \par - Controllability & - Availability \par - Accuracy \par - Explainability & & - AI alignment \cite{ouyang2022training}\par - Method for providing explainability while concealing model information \\ \hline

    \multirow{3}{*}{Military use} & Using AI for military purposes &  - Controllability& - Availability\par - Accuracy \par - Explainability & & - AI alignment \cite{ouyang2022training}\\ 
    \hline

    \multirow{19}{*}{Privacy} & Privacy violation due to the leakage of personal information from AI &  - Confidentiality\par - Integrity &  &  & - DP \cite{abadi2016deep}\par - Federated learning \cite{wang2020federated}\par - AI alignment \cite{ouyang2022training}\par - Machine unlearning \cite{tramer2022detecting} \par - Encryption technology \cite{juvekar2018gazelle}\par - Anonymization technology \cite{slijepvcevic2021k}\\  \cline{2-6}

    & Using fragmented information and AI to make inferences and identify personal information &  & & - Attacks that use AI to analyze information collected from social media to identify individuals & - Anonymization technology \cite{slijepvcevic2021k}\par - DP \cite{abadi2016deep}\par - Federated learning \cite{wang2020federated}\par - Machine unlearning \cite{tramer2022detecting} \par -  Encryption technology \cite{juvekar2018gazelle} \\ \cline{2-6}

    & Inferring a person's character or personal preferences from their facial expressions &  & & - Attacks that use AI to analyze images to infer personal information & - Anonymization technology \cite{slijepvcevic2021k}\par - DP \cite{abadi2016deep}\par - Federated learning \cite{wang2020federated}\par - Machine unlearning \cite{tramer2022detecting} \par - Encryption technology \cite{juvekar2018gazelle} \\ \cline{2-6}

    & Using AI-generated emails or audio to prompt the input of confidential information and steal it &  & - Availability \par - Accuracy & - Social engineering attack \cite{schmitt2024digital} & - Anonymization technology \cite{slijepvcevic2021k}\par - DP \cite{abadi2016deep}\par - Federated learning \cite{wang2020federated}\par - Machine unlearning \cite{tramer2022detecting} \par - Encryption technology \cite{juvekar2018gazelle} \\ 
    \hline

    \multirow{6}{*}{Disinformation} & Creating disinformation using AI & - Controllability\par - Integrity & - Availability\par - Accuracy\par & - Deepfake\par - Social engineering attack \cite{schmitt2024digital} & - AI alignment \cite{ouyang2022training}\par - Watermarking for generative AI \cite{kirchenbauer2023watermark} \par - Encryption technology \cite{juvekar2018gazelle}\par - Identification of AI-generated output\par - Detection of disinformation\par - Deepfake detection \cite{wang2020cnn}\\ 
    \hline

    \multirow{3}{*}{Copyright and authorship} & Violation of copyright and authorship by AI-generated similar content &  - Integrity\par - Controllability\par & & - Plagiarism & - Defensive methods for integrity and plagiarism \\ 
    \hline

    \multirow{2}{*}{Human-centric principle} & Improperly manipulating the decision-making of non-consumers by AI & & &- Disinformation  & - Defensive methods for integrity and disinformation\par - Human in the loop\\ 
    \hline

    \multirow{2}{*}{Ethics} & Unethical output or actions by AI & - Integrity & & - Jailbreak & - Education and follow-up\par - AI alignment \cite{ouyang2022training}\\ 
    \hline

    \multirow{4}{*}{Physical impact} & Physical harm to non-consumers caused by AI &  - Integrity \par - Accuracy\par - Controllability\par - Safety & & - Military use & - Defensive methods for integrity \\ \hline

     \multirow{5}{*}{Psychological impact} & Psychological harm to non-consumers caused by AI &  - Confidentiality\par - Controllability\par - Safety\par - Output fairness\par & & - Disinformation\par - Military use & - Defensive methods for integrity\par - Defensive methods for availability\par - Defensive methods for disinformation \\ \hline

    \multirow{2}{*}{Financial impact} & Financial harm to non-consumers caused by AI & - Safety & & - Cyber attacks\par - Disinformation & - Defensive methods for integrity\par - Defensive methods for disinformation \\ \hline

    \end{tabular}

}

\end{table*}

\smallskip
\noindent\textbf{Society.}
Table~\ref{tab:EIA_society_provider} shows the AI security map for negative impacts on society and AI system providers in the EIA.
We identified nine negative impacts on society and eight related elements. 
It is found that the compromise of ISA elements is closely linked to impacts on society as well. 
In particular, impacts on healthcare and critical infrastructure are more likely to occur, due to the large number of relevant ISA elements involved. 
Given the anticipated widespread utilization of AI in society, it is important to consider potential impacts on these fields and to develop appropriate countermeasures. Furthermore, the misuse of ISA elements can also lead to the spread of disinformation, cyber attacks, and violations of laws and regulations. 
For example, the dissemination of disinformation through deepfakes is a representative negative impact, resulting from both the availability and the abuse of deepfake technology and its accuracy.

\smallskip
\noindent\textbf{AI system providers.}
We identified two negative impacts on AI system providers and two related elements. We believe that impacts on AI system providers can result from the compromise of any of the ISA elements. Furthermore, negative impacts caused by misuse can lead to reputational and financial consequences for AI system providers. As the number of companies and individuals developing AI-based systems increases, it is important that these risks are properly recognized.

\begin{table*}[t]
\centering
\caption{AI Security Map for negative impacts on society and AI system providers in the EIA.}
\label{tab:EIA_society_provider}

\huge
\scalebox{0.42}{

    \begin{tabular}{|c|p{0.5\textwidth}|p{0.35\textwidth}|p{0.2\textwidth}|p{0.45\textwidth}|p{0.45\textwidth}|} 
    \hline
    
    \multirow{2}{*}{\textbf{Elements}} & \multirow{2}{*}{\textbf{Negative impacts}} & \multicolumn{2}{c|}{\textbf{Related elements in the ISA}} &  \textbf{Causal factors or } & \multirow{1}{*}{\textbf{Defensive methods or}} \\ \cline{3-4}
    &  & \textbf{Compromise} & \textbf{Abuse} & \textbf{related elements in the EIA} & \textbf{countermeasures} \\ \hline
    
    \multicolumn{6}{|c|}{\textbf{Society}}  \\ \hline

    \multirow{3}{*}{Cyber attack} & Using AI for cyber attacks & - Confidentiality \par - Controllability & - Availability \par - Accuracy \par - explainability & & - AI alignment \cite{ouyang2022training}\par - Method for providing explainability while concealing model information \\ \hline

    \multirow{3}{*}{Military use} & Using AI for military purposes &  - Controllability& - Availability\par - Accuracy \par - Explainability & & - AI alignment \cite{ouyang2022training}\\ 
    \hline

    \multirow{6}{*}{Disinformation} & Creating disinformation using AI & - Controllability\par & - Availability\par - Accuracy\par & - Deepfake\par - Social engineering attack \cite{schmitt2024digital} & - AI alignment \cite{ouyang2022training}\par - Watermarking for generative AI \cite{kirchenbauer2023watermark}\par - Encryption technology \cite{juvekar2018gazelle}\par - Identification of AI-generated output\par - Detection of disinformation\par - Deepfake detection \cite{wang2020cnn}\\ 
    \hline

    \multirow{4}{*}{Compliance with} & Using AI for purposes that violate the law &  - Confidentiality\par - Controllability & - Availability\par - Accuracy& & - AI alignment \cite{ouyang2022training}\par - AI access control \\ \cline{2-6}
    
    \multirow{2}{*}{laws and regulations} & Actions that violate the law by AI &  - Integrity \par - Accuracy\par - Controllability & & & - AI alignment \cite{ouyang2022training}\par - AI access control \\ 
    \hline

    \multirow{2}{*}{Ethics} & Unethical output or actions by AI & - Integrity & & - Jailbreak & - Education and follow-up\par - AI alignment \cite{ouyang2022training}\\ 
    \hline

    \multirow{4}{*}{Economy} & AI negatively impacting the economy & - Safety\par - Accuracy\par & & - Cyber attack\par - Military use\par - Disinformation\par - Human-centric principle & - Defensive methods for integrity\par - Defensive methods for disinformation \\ 
    \hline
    
    \multirow{8}{*}{Medical care} & Negative impact on medical care caused by AI & - Integrity \par - Availability\par - Accuracy\par - Controllability \par - Safety \par - Output fairness\par - Explainability\par - Trustworthiness & & - Human-centric principle & - Defensive methods for integrity \par - Defensive methods for availability \\ 
    \hline

    \multirow{8}{*}{Critical infrastructure} & Negative impact on critical infrastructure caused by AI & - Integrity \par - Availability \par - Accuracy\par - Controllability\par - Safety\par - Output fairness\par - Explainability \par - Trustworthiness & & - Human-centric principle & - Defensive methods for integrity \par - Defensive methods for availability \\ 
    \hline

    \multicolumn{6}{|c|}{\textbf{AI system providers}} \\ \hline
    
    
    \multirow{3}{*}{Reputation} & The decline in the reputation of AI system providers & - All the elements in the ISA & & - Cyber attack \par - Military use \par - Compliance with laws and regulations & - Defensive methods for confidentiality\par - Defensive methods for integrity\par - Defensive methods for availability \\ 
    \hline
    \multirow{2}{*}{Financial impact} & Financial harm to AI system providers caused by AI & - All the elements in the ISA & & - Cyber attack\par - Compliance with laws and regulations & - Defensive methods for integrity \par - Defensive methods for availability \\ 
    \hline

    \end{tabular}

}

\end{table*}

\begin{boxtakeaway}
Most negative impacts on \textbf{consumers} tend to be linked to the compromise of AI elements in the ISA.
However, some can result directly from targeted attacks, such as social engineering \cite{schmitt2024digital} or poisoning \cite{carlini2024poisoning,biggio2012poisoning}, leading to privacy and misinformation issues for consumers. 
\textbf{Non-consumers} can also experience negative impacts, due to the abuse of AI elements by attackers, especially in the form of privacy violations and the spread of misinformation. 
\textbf{Society} can incur significant impacts, particularly concerning critical infrastructure and medical care.
AI misuse can cause widespread harm such as disinformation, cyber attacks, and legal violations. 
For \textbf{AI system providers}, negative impacts can lead to reputation and financial damage, and these risks increase as more entities develop AI systems. 
\end{boxtakeaway}

\subsection{Relationships between Two Aspects}
Basically, we assume that the compromise of any element in the ISA have impacts on the elements of the EIA. 
In a nutshell, we consider that negative impacts on individuals and society may occur in a chain reaction from the compromise of information systems in many cases.
By referring elements categorized into the two aspects, we can reveal how negative impacts affect individuals and society.
We found that there are two types of chains leading to negative impacts on individuals and society.
The first type is the case where the compromise of AI in information systems has a direct influence on individuals and society. 
In this case, negative impacts in the ISA are directly related to those on individuals and society.
The second type is the case where negative impacts affect individuals and society indirectly, through multiple negative impacts in the EIA.
In what follows, we describe each of these chains in detail.

\smallskip
\noindent\textbf{Direct chain of impacts.} 
Negative impacts caused by the compromise of elements within the ISA often directly affect individuals and society. 
A typical example is privacy. 
It is easy to imagine that when ``confidentiality'' is compromised, ``privacy'' is also breached. 
In the context of AI security, where AI can retain knowledge about vast amounts of data, the impact on non-consumers can be significant. 
Therefore, from the perspective of promoting AI utilization, technologies that ensure confidentiality and privacy protection are considered important. 
Furthermore, as shown in Table~\ref{tab:ISA_CIA}, negative impacts caused by the integrity violation are related to many external elements, such as human-centered principles, medical care, and critical infrastructure. 
As a result, it has been found that a compromise of integrity, in particular, can have a significant social impact.  
As with the integrity violation, negative impacts resulting from the availability breach directly relate to many elements in the EIA.

\smallskip
\noindent\textbf{Indirect chain of impacts.} 
In addition to cases where individuals or society are directly affected, there are also cases where multiple impacts occur in a chain, ultimately affecting individuals or society. 
For example, as shown in Table~\ref{tab:ISA_others}, if the ``integrity'' of an LLM is compromised through a prompt injection attack, ``controllability'' is first affected. 
Once controllability is compromised, attackers can make the LLM generate ``disinformation'' as they intend. 
If this disinformation spreads, it can influence human decision-making, thereby causing negative effects in terms of ``human-centered principles'' as shown in Table~\ref{tab:EIA_non-consumer}. 
As a result, non-consumers who see this disinformation may also be affected.  
Furthermore, the misuse of AI systems can occur by taking advantage of the fact that elements in the ISA are satisfied.
This can also have a cascading effect on individuals and society. 
For instance, cyber attacks through malware generation using LLMs exploit ``accuracy'' and ``availability'', leading to ``financial impacts'' on non-consumers and negative ``economic'' effects on society. 
In particular, such ripple effects tend to influence non-consumers who do not use AI at all. 
It is important to consider how to prevent negative impacts on non-consumers and society from such misuse, while also ensuring convenience for consumers.

\begin{boxtakeaway}
There are the two types of chains in terms of impacts on people and society.
In particular, the misuse of AI can occur by exploiting the fulfilled elements of the ISA.
Therefore, in addition to AI-specific defense methods, considering alternative defense approaches with zero trust security helps mitigate negative impacts on individuals and society.
\end{boxtakeaway}

\section{Recommendations} \label{sec:recommend}
We present some recommendations in response to the current landscape of AI security and the holistic organization in our work.
We consider recommendations from the two perspectives, namely fundamental research and applied research.

\subsection{Fundamental Research on AI Security}
As shown above, the impacts of AI and ML extend beyond traditional security concepts and the scope of privacy. 
Hence, we encourage researchers working on future studies to holistically analyze what kinds of impacts may occur with holistic overviews such as our map when they consider new attacks or countermeasures. 
By doing this, it may help them devise more innovative research directions and more practical methods, thereby enhancing AI security.
We hope that our AI security map will contribute to the promotion of fundamental research, as well as the discovery of new elements, negative impacts, and security targets.

Furthermore, we recommend that more active efforts be made to pursue holistic organization as we have proposed.
In what follows, we present several \textbf{open problems} and future directions in terms of the holistic organization of AI security for researchers.

\smallskip
\noindent\textbf{Appropriate Granularity in Defining security targets.}
At this stage, security targets in our map are defined in relatively broad categories. 
However, in reality, a more diverse range of stakeholders should be considered. 
For example, since ``consumers'' may include both decision-makers and developers, each of whom may require different types of information. 
Therefore, it is necessary to establish a more detailed definition of stakeholders in future work.

\smallskip
\noindent\textbf{Quantitative Assessment of Impacts.}
Our map helps understand a holistic overview of AI security.
However, the degree of each impact is not clear at this stage.
Currently, only the relationships among various factors are identified, and quantitative risk assessment has not yet been addressed. 
To enhance the informativeness of the analysis with the AI security map, it is important to indicate the extent of risk associated with each negative impact in future.
Since the level of risk may vary depending on the security targets and the chain of impacts, it is important to develop quantification methods that take these factors into account.

\smallskip
\noindent\textbf{Mapping New Domains.}
It is also necessary to consider new domains such as the security of agentic AI.
Additionally, there is room for discussion regarding the positioning of methods such as safety verification and red teaming, which are not direct countermeasures. 
While these techniques are important for proactively identifying vulnerabilities and assessing risks in AI systems, their classification should be carefully considered, as they are more akin to preventive measures.
Due to the extremely rapid emergence of new AI technologies, it is also necessary to establish mechanisms and methodologies for efficiently keeping up with these advancements and promptly reflecting them in the holistic organization.

\smallskip
\noindent\textbf{Automation of the Definition and Classification of Negative Impacts and Elements.}
At this stage, elements and negative impacts are identified and classified manually by referencing multiple papers and guidelines. 
Given the rapid emergence of new AI technologies, it may become necessary to automatically define and classify elements in response to the swift evolution of AI. 
This challenge is also relevant to conventional survey papers because it is likely to be difficult to comprehensively cover the vast amount of information manually without omissions. 
Thus, the development of systematization technologies utilizing LLMs and AI is interesting and essential in future.

\subsection{Applied Research on AI Security}
In addition to fundamental research, it is also important to consider how the holistic overview can be utilized and further developed from the perspective of applied research.
For people other than AI security researchers and experts, it is important to be aware of the potential attacks and risks associated with the utilization of AI.
In particular, business stakeholders and decision-makers who consider incorporating AI into their services or products may be interested in negative impacts entailed by introducing AI.
For example, one possible application is obtaining relevant negative impacts by inputting a news article about AI security into an LLM with our map.
By doing this, they can gain insights into relevant negative impacts without expertise.
We actually conducted a basic trial of this use case.
To this end, we input a news article about AI being misused for cyber attacks into GPT-4.1 \cite{openai-gpt4-1}.
The prompt used for this use case is shown in Appendix~\ref{prompt_example}.
As a result, the LLM outputs negative impacts regarding four elements (integrity, availability, trustworthiness, and accuracy) in the ISA and two elements (cyber attacks and human-centric principle) in the EIA.
While elements related to cyber attacks can be easily inferred from the content of the news, ``human-centric principle'' is not as readily apparent. 
Therefore, this result demonstrates that valuable information in the EIA can be obtained.
Such a use case is also expected to be highly beneficial for consumers.

Another application of our map is that AI system providers or developers use our map in order to to understand negative impacts on AI within information systems.
It is essential for them to be aware of the potential negative impacts in the EIA when designing and implementing information systems.
Depending on the service, the primary concern required to minimize negative impacts on end users is different.
Thus, such a use case is helpful in determining which element in the ISA should be developed with greater robustness. 
Being informed of these possible impacts in advance enables AI system providers or developers to anticipate adverse outcomes and design systems more efficiently and effectively.

By implementing the above use cases, it is possible to facilitate a better understanding of the complex technologies involved in AI security, the potential negative impacts on each security target, and the relationships among various elements.
Since the above use cases are just examples, other useful application or practical methods should be devised in future.

\section{Conclusion}
In this paper, we have developed the AI security map to provide a holistic overview of AI security. 
The map identifies key elements that AI systems should fulfill, as well as factors that influence individuals and society, from both the perspective of the ISA and EIA.
Furthermore, we holistically organized the relationships between AI elements and external factors in the EIA, which had not previously been addressed. 
In short, our map helps clarify how damage to AI-based information systems can affect people and society. 
We also categorized the specific negative impacts associated with each element, along with their causes including attacks, causal factors, and countermeasures. 
By distinguishing whether these causes originate from the compromise or misuse of elements in the ISA, we clarify how AI can lead to various consequences.
Through the insights and recommendations derived from our map, we discuss the value of holistically organizing AI security knowledge and technology. 
We hope that our work will serve as an important foundation for researchers and a wide range of stakeholders, facilitating the collection and understanding of AI security information in this complex field, and promoting further discovery and research on new elements and negative impact targets.

\begin{acks}
This work was supported by JST K Program, Japan Grant Number JPMJKP24C4.
\end{acks}

\bibliographystyle{ACM-Reference-Format}
\bibliography{reference}


\begin{thebibliography}{62}


\ifx \showCODEN    \undefined \def \showCODEN     #1{\unskip}     \fi
\ifx \showISBNx    \undefined \def \showISBNx     #1{\unskip}     \fi
\ifx \showISBNxiii \undefined \def \showISBNxiii  #1{\unskip}     \fi
\ifx \showISSN     \undefined \def \showISSN      #1{\unskip}     \fi
\ifx \showLCCN     \undefined \def \showLCCN      #1{\unskip}     \fi
\ifx \shownote     \undefined \def \shownote      #1{#1}          \fi
\ifx \showarticletitle \undefined \def \showarticletitle #1{#1}   \fi
\ifx \showURL      \undefined \def \showURL       {\relax}        \fi
\providecommand\bibfield[2]{#2}
\providecommand\bibinfo[2]{#2}
\providecommand\natexlab[1]{#1}
\providecommand\showeprint[2][]{arXiv:#2}

\bibitem[Abadi et~al\mbox{.}(2016)]%
        {abadi2016deep}
\bibfield{author}{\bibinfo{person}{Martin Abadi}, \bibinfo{person}{Andy Chu}, \bibinfo{person}{Ian Goodfellow}, \bibinfo{person}{H~Brendan McMahan}, \bibinfo{person}{Ilya Mironov}, \bibinfo{person}{Kunal Talwar}, {and} \bibinfo{person}{Li Zhang}.} \bibinfo{year}{2016}\natexlab{}.
\newblock \showarticletitle{Deep learning with differential privacy}. In \bibinfo{booktitle}{\emph{Proceedings of the 2016 ACM SIGSAC conference on computer and communications security}}. \bibinfo{pages}{308--318}.
\newblock


\bibitem[A{\"\i}meur et~al\mbox{.}(2023)]%
        {aimeur2023fake}
\bibfield{author}{\bibinfo{person}{Esma A{\"\i}meur}, \bibinfo{person}{Sabrine Amri}, {and} \bibinfo{person}{Gilles Brassard}.} \bibinfo{year}{2023}\natexlab{}.
\newblock \showarticletitle{Fake news, disinformation and misinformation in social media: a review}.
\newblock \bibinfo{journal}{\emph{Social Network Analysis and Mining}} \bibinfo{volume}{13}, \bibinfo{number}{1} (\bibinfo{year}{2023}), \bibinfo{pages}{30}.
\newblock


\bibitem[Asai et~al\mbox{.}(2023)]%
        {asai2023self}
\bibfield{author}{\bibinfo{person}{Akari Asai}, \bibinfo{person}{Zeqiu Wu}, \bibinfo{person}{Yizhong Wang}, \bibinfo{person}{Avirup Sil}, {and} \bibinfo{person}{Hannaneh Hajishirzi}.} \bibinfo{year}{2023}\natexlab{}.
\newblock \showarticletitle{Self-rag: Learning to retrieve, generate, and critique through self-reflection}. In \bibinfo{booktitle}{\emph{The Twelfth International Conference on Learning Representations}}.
\newblock


\bibitem[Biggio et~al\mbox{.}(2012)]%
        {biggio2012poisoning}
\bibfield{author}{\bibinfo{person}{Battista Biggio}, \bibinfo{person}{Blaine Nelson}, {and} \bibinfo{person}{Pavel Laskov}.} \bibinfo{year}{2012}\natexlab{}.
\newblock \showarticletitle{Poisoning attacks against support vector machines}.
\newblock \bibinfo{journal}{\emph{arXiv preprint arXiv:1206.6389}} (\bibinfo{year}{2012}).
\newblock


\bibitem[Carlini et~al\mbox{.}(2024)]%
        {carlini2024poisoning}
\bibfield{author}{\bibinfo{person}{Nicholas Carlini}, \bibinfo{person}{Matthew Jagielski}, \bibinfo{person}{Christopher~A Choquette-Choo}, \bibinfo{person}{Daniel Paleka}, \bibinfo{person}{Will Pearce}, \bibinfo{person}{Hyrum Anderson}, \bibinfo{person}{Andreas Terzis}, \bibinfo{person}{Kurt Thomas}, {and} \bibinfo{person}{Florian Tram{\`e}r}.} \bibinfo{year}{2024}\natexlab{}.
\newblock \showarticletitle{Poisoning web-scale training datasets is practical}. In \bibinfo{booktitle}{\emph{2024 IEEE Symposium on Security and Privacy (SP)}}. IEEE, \bibinfo{pages}{407--425}.
\newblock


\bibitem[Castagnaro et~al\mbox{.}(2024)]%
        {castagnaro2024offensive}
\bibfield{author}{\bibinfo{person}{Alberto Castagnaro}, \bibinfo{person}{Mauro Conti}, {and} \bibinfo{person}{Luca Pajola}.} \bibinfo{year}{2024}\natexlab{}.
\newblock \showarticletitle{Offensive AI: Enhancing Directory Brute-forcing Attack with the Use of Language Models}. In \bibinfo{booktitle}{\emph{Proceedings of the 2024 Workshop on Artificial Intelligence and Security}}. \bibinfo{pages}{184--195}.
\newblock


\bibitem[Chang et~al\mbox{.}(2024)]%
        {chang2024sok}
\bibfield{author}{\bibinfo{person}{Yuanhaur Chang}, \bibinfo{person}{Han Liu}, \bibinfo{person}{Evin Jaff}, \bibinfo{person}{Chenyang Lu}, {and} \bibinfo{person}{Ning Zhang}.} \bibinfo{year}{2024}\natexlab{}.
\newblock \showarticletitle{SoK: Security and Privacy Risks of Medical AI}.
\newblock \bibinfo{journal}{\emph{arXiv preprint arXiv:2409.07415}} (\bibinfo{year}{2024}).
\newblock


\bibitem[Chen and Babar(2024)]%
        {chen2024security}
\bibfield{author}{\bibinfo{person}{Huaming Chen} {and} \bibinfo{person}{M~Ali Babar}.} \bibinfo{year}{2024}\natexlab{}.
\newblock \showarticletitle{Security for machine learning-based software systems: A survey of threats, practices, and challenges}.
\newblock \bibinfo{journal}{\emph{Comput. Surveys}} \bibinfo{volume}{56}, \bibinfo{number}{6} (\bibinfo{year}{2024}), \bibinfo{pages}{1--38}.
\newblock


\bibitem[Dibbo(2023)]%
        {dibbo2023sok}
\bibfield{author}{\bibinfo{person}{Sayanton~V Dibbo}.} \bibinfo{year}{2023}\natexlab{}.
\newblock \showarticletitle{Sok: Model inversion attack landscape: Taxonomy, challenges, and future roadmap}. In \bibinfo{booktitle}{\emph{2023 IEEE 36th Computer Security Foundations Symposium (CSF)}}. IEEE, \bibinfo{pages}{439--456}.
\newblock


\bibitem[Doan et~al\mbox{.}(2020)]%
        {doan2020februus}
\bibfield{author}{\bibinfo{person}{Bao~Gia Doan}, \bibinfo{person}{Ehsan Abbasnejad}, {and} \bibinfo{person}{Damith~C Ranasinghe}.} \bibinfo{year}{2020}\natexlab{}.
\newblock \showarticletitle{Februus: Input purification defense against trojan attacks on deep neural network systems}. In \bibinfo{booktitle}{\emph{Proceedings of the 36th Annual Computer Security Applications Conference}}. \bibinfo{pages}{897--912}.
\newblock


\bibitem[Dwivedi et~al\mbox{.}(2023)]%
        {dwivedi2023explainable}
\bibfield{author}{\bibinfo{person}{Rudresh Dwivedi}, \bibinfo{person}{Devam Dave}, \bibinfo{person}{Het Naik}, \bibinfo{person}{Smiti Singhal}, \bibinfo{person}{Rana Omer}, \bibinfo{person}{Pankesh Patel}, \bibinfo{person}{Bin Qian}, \bibinfo{person}{Zhenyu Wen}, \bibinfo{person}{Tejal Shah}, \bibinfo{person}{Graham Morgan}, {et~al\mbox{.}}} \bibinfo{year}{2023}\natexlab{}.
\newblock \showarticletitle{Explainable AI (XAI): Core ideas, techniques, and solutions}.
\newblock \bibinfo{journal}{\emph{Comput. Surveys}} \bibinfo{volume}{55}, \bibinfo{number}{9} (\bibinfo{year}{2023}), \bibinfo{pages}{1--33}.
\newblock


\bibitem[Golda et~al\mbox{.}(2024)]%
        {golda2024privacy}
\bibfield{author}{\bibinfo{person}{Abenezer Golda}, \bibinfo{person}{Kidus Mekonen}, \bibinfo{person}{Amit Pandey}, \bibinfo{person}{Anushka Singh}, \bibinfo{person}{Vikas Hassija}, \bibinfo{person}{Vinay Chamola}, {and} \bibinfo{person}{Biplab Sikdar}.} \bibinfo{year}{2024}\natexlab{}.
\newblock \showarticletitle{Privacy and security concerns in generative AI: a comprehensive survey}.
\newblock \bibinfo{journal}{\emph{IEEE Access}} (\bibinfo{year}{2024}).
\newblock


\bibitem[Greshake et~al\mbox{.}(2023)]%
        {greshake2023not}
\bibfield{author}{\bibinfo{person}{Kai Greshake}, \bibinfo{person}{Sahar Abdelnabi}, \bibinfo{person}{Shailesh Mishra}, \bibinfo{person}{Christoph Endres}, \bibinfo{person}{Thorsten Holz}, {and} \bibinfo{person}{Mario Fritz}.} \bibinfo{year}{2023}\natexlab{}.
\newblock \showarticletitle{Not what you've signed up for: Compromising real-world llm-integrated applications with indirect prompt injection}. In \bibinfo{booktitle}{\emph{Proceedings of the 16th ACM Workshop on Artificial Intelligence and Security}}. \bibinfo{pages}{79--90}.
\newblock


\bibitem[Gu et~al\mbox{.}(2017)]%
        {gu2017badnets}
\bibfield{author}{\bibinfo{person}{Tianyu Gu}, \bibinfo{person}{Brendan Dolan-Gavitt}, {and} \bibinfo{person}{Siddharth Garg}.} \bibinfo{year}{2017}\natexlab{}.
\newblock \showarticletitle{Badnets: Identifying vulnerabilities in the machine learning model supply chain}.
\newblock \bibinfo{journal}{\emph{arXiv preprint arXiv:1708.06733}} (\bibinfo{year}{2017}).
\newblock


\bibitem[Hu et~al\mbox{.}(2021)]%
        {hu2021artificial}
\bibfield{author}{\bibinfo{person}{Yupeng Hu}, \bibinfo{person}{Wenxin Kuang}, \bibinfo{person}{Zheng Qin}, \bibinfo{person}{Kenli Li}, \bibinfo{person}{Jiliang Zhang}, \bibinfo{person}{Yansong Gao}, \bibinfo{person}{Wenjia Li}, {and} \bibinfo{person}{Keqin Li}.} \bibinfo{year}{2021}\natexlab{}.
\newblock \showarticletitle{Artificial intelligence security: Threats and countermeasures}.
\newblock \bibinfo{journal}{\emph{ACM Computing Surveys (CSUR)}} \bibinfo{volume}{55}, \bibinfo{number}{1} (\bibinfo{year}{2021}), \bibinfo{pages}{1--36}.
\newblock


\bibitem[Hu et~al\mbox{.}(2024)]%
        {hu2024toxicity}
\bibfield{author}{\bibinfo{person}{Zhanhao Hu}, \bibinfo{person}{Julien Piet}, \bibinfo{person}{Geng Zhao}, \bibinfo{person}{Jiantao Jiao}, {and} \bibinfo{person}{David Wagner}.} \bibinfo{year}{2024}\natexlab{}.
\newblock \showarticletitle{Toxicity detection for free}.
\newblock \bibinfo{journal}{\emph{arXiv preprint arXiv:2405.18822}} (\bibinfo{year}{2024}).
\newblock


\bibitem[Hui et~al\mbox{.}(2024)]%
        {hui2024pleak}
\bibfield{author}{\bibinfo{person}{Bo Hui}, \bibinfo{person}{Haolin Yuan}, \bibinfo{person}{Neil Gong}, \bibinfo{person}{Philippe Burlina}, {and} \bibinfo{person}{Yinzhi Cao}.} \bibinfo{year}{2024}\natexlab{}.
\newblock \showarticletitle{Pleak: Prompt leaking attacks against large language model applications}. In \bibinfo{booktitle}{\emph{Proceedings of the 2024 on ACM SIGSAC Conference on Computer and Communications Security}}. \bibinfo{pages}{3600--3614}.
\newblock


\bibitem[Juvekar et~al\mbox{.}(2018)]%
        {juvekar2018gazelle}
\bibfield{author}{\bibinfo{person}{Chiraag Juvekar}, \bibinfo{person}{Vinod Vaikuntanathan}, {and} \bibinfo{person}{Anantha Chandrakasan}.} \bibinfo{year}{2018}\natexlab{}.
\newblock \showarticletitle{$\{$GAZELLE$\}$: A low latency framework for secure neural network inference}. In \bibinfo{booktitle}{\emph{27th USENIX security symposium (USENIX security 18)}}. \bibinfo{pages}{1651--1669}.
\newblock


\bibitem[Kheya et~al\mbox{.}(2024)]%
        {kheya2024pursuit}
\bibfield{author}{\bibinfo{person}{Tahsin~Alamgir Kheya}, \bibinfo{person}{Mohamed~Reda Bouadjenek}, {and} \bibinfo{person}{Sunil Aryal}.} \bibinfo{year}{2024}\natexlab{}.
\newblock \showarticletitle{The pursuit of fairness in artificial intelligence models: A survey}.
\newblock \bibinfo{journal}{\emph{arXiv preprint arXiv:2403.17333}} (\bibinfo{year}{2024}).
\newblock


\bibitem[Kirchenbauer et~al\mbox{.}(2023)]%
        {kirchenbauer2023watermark}
\bibfield{author}{\bibinfo{person}{John Kirchenbauer}, \bibinfo{person}{Jonas Geiping}, \bibinfo{person}{Yuxin Wen}, \bibinfo{person}{Jonathan Katz}, \bibinfo{person}{Ian Miers}, {and} \bibinfo{person}{Tom Goldstein}.} \bibinfo{year}{2023}\natexlab{}.
\newblock \showarticletitle{A watermark for large language models}. In \bibinfo{booktitle}{\emph{International Conference on Machine Learning}}. PMLR, \bibinfo{pages}{17061--17084}.
\newblock


\bibitem[Kiribuchi et~al\mbox{.}(2025)]%
        {kiribuchi2025securingaisystemsguide}
\bibfield{author}{\bibinfo{person}{Naoto Kiribuchi}, \bibinfo{person}{Kengo Zenitani}, {and} \bibinfo{person}{Takayuki Semitsu}.} \bibinfo{year}{2025}\natexlab{}.
\newblock \bibinfo{title}{Securing AI Systems: A Guide to Known Attacks and Impacts}.
\newblock
\showeprint[arxiv]{2506.23296}~[cs.CR]
\urldef\tempurl%
\url{https://arxiv.org/abs/2506.23296}
\showURL{%
\tempurl}


\bibitem[Lee et~al\mbox{.}(2024)]%
        {lee2024deepfakes}
\bibfield{author}{\bibinfo{person}{Hao-Ping Lee}, \bibinfo{person}{Yu-Ju Yang}, \bibinfo{person}{Thomas~Serban Von~Davier}, \bibinfo{person}{Jodi Forlizzi}, {and} \bibinfo{person}{Sauvik Das}.} \bibinfo{year}{2024}\natexlab{}.
\newblock \showarticletitle{Deepfakes, phrenology, surveillance, and more! a taxonomy of ai privacy risks}. In \bibinfo{booktitle}{\emph{Proceedings of the 2024 CHI Conference on Human Factors in Computing Systems}}. \bibinfo{pages}{1--19}.
\newblock


\bibitem[Li et~al\mbox{.}(2023)]%
        {li2023sok}
\bibfield{author}{\bibinfo{person}{Linyi Li}, \bibinfo{person}{Tao Xie}, {and} \bibinfo{person}{Bo Li}.} \bibinfo{year}{2023}\natexlab{}.
\newblock \showarticletitle{Sok: Certified robustness for deep neural networks}. In \bibinfo{booktitle}{\emph{2023 IEEE symposium on security and privacy (SP)}}. IEEE, \bibinfo{pages}{1289--1310}.
\newblock


\bibitem[Lu et~al\mbox{.}(2024)]%
        {lu2024autojailbreak}
\bibfield{author}{\bibinfo{person}{Lin Lu}, \bibinfo{person}{Hai Yan}, \bibinfo{person}{Zenghui Yuan}, \bibinfo{person}{Jiawen Shi}, \bibinfo{person}{Wenqi Wei}, \bibinfo{person}{Pin-Yu Chen}, {and} \bibinfo{person}{Pan Zhou}.} \bibinfo{year}{2024}\natexlab{}.
\newblock \showarticletitle{Autojailbreak: Exploring jailbreak attacks and defenses through a dependency lens}.
\newblock \bibinfo{journal}{\emph{arXiv preprint arXiv:2406.03805}} (\bibinfo{year}{2024}).
\newblock


\bibitem[Madry et~al\mbox{.}(2017)]%
        {madry2017towards}
\bibfield{author}{\bibinfo{person}{Aleksander Madry}, \bibinfo{person}{Aleksandar Makelov}, \bibinfo{person}{Ludwig Schmidt}, \bibinfo{person}{Dimitris Tsipras}, {and} \bibinfo{person}{Adrian Vladu}.} \bibinfo{year}{2017}\natexlab{}.
\newblock \showarticletitle{Towards deep learning models resistant to adversarial attacks}.
\newblock \bibinfo{journal}{\emph{arXiv preprint arXiv:1706.06083}} (\bibinfo{year}{2017}).
\newblock


\bibitem[Mersha et~al\mbox{.}(2024)]%
        {mersha2024explainable}
\bibfield{author}{\bibinfo{person}{Melkamu Mersha}, \bibinfo{person}{Khang Lam}, \bibinfo{person}{Joseph Wood}, \bibinfo{person}{Ali AlShami}, {and} \bibinfo{person}{Jugal Kalita}.} \bibinfo{year}{2024}\natexlab{}.
\newblock \showarticletitle{Explainable artificial intelligence: A survey of needs, techniques, applications, and future direction}.
\newblock \bibinfo{journal}{\emph{Neurocomputing}} (\bibinfo{year}{2024}), \bibinfo{pages}{128111}.
\newblock


\bibitem[Mothukuri et~al\mbox{.}(2021)]%
        {mothukuri2021survey}
\bibfield{author}{\bibinfo{person}{Viraaji Mothukuri}, \bibinfo{person}{Reza~M Parizi}, \bibinfo{person}{Seyedamin Pouriyeh}, \bibinfo{person}{Yan Huang}, \bibinfo{person}{Ali Dehghantanha}, {and} \bibinfo{person}{Gautam Srivastava}.} \bibinfo{year}{2021}\natexlab{}.
\newblock \showarticletitle{A survey on security and privacy of federated learning}.
\newblock \bibinfo{journal}{\emph{Future Generation Computer Systems}}  \bibinfo{volume}{115} (\bibinfo{year}{2021}), \bibinfo{pages}{619--640}.
\newblock


\bibitem[Nazary et~al\mbox{.}(2025)]%
        {nazary2025poison}
\bibfield{author}{\bibinfo{person}{Fatemeh Nazary}, \bibinfo{person}{Yashar Deldjoo}, {and} \bibinfo{person}{Tommaso~di Noia}.} \bibinfo{year}{2025}\natexlab{}.
\newblock \showarticletitle{Poison-rag: Adversarial data poisoning attacks on retrieval-augmented generation in recommender systems}. In \bibinfo{booktitle}{\emph{European Conference on Information Retrieval}}. Springer, \bibinfo{pages}{239--251}.
\newblock


\bibitem[Noppel and Wressnegger(2024)]%
        {noppel2024sok}
\bibfield{author}{\bibinfo{person}{Maximilian Noppel} {and} \bibinfo{person}{Christian Wressnegger}.} \bibinfo{year}{2024}\natexlab{}.
\newblock \showarticletitle{SoK: Explainable machine learning in adversarial environments}. In \bibinfo{booktitle}{\emph{2024 IEEE Symposium on Security and Privacy (SP)}}. IEEE, \bibinfo{pages}{2441--2459}.
\newblock


\bibitem[Northcutt et~al\mbox{.}(2021)]%
        {northcutt2021pervasive}
\bibfield{author}{\bibinfo{person}{Curtis~G Northcutt}, \bibinfo{person}{Anish Athalye}, {and} \bibinfo{person}{Jonas Mueller}.} \bibinfo{year}{2021}\natexlab{}.
\newblock \showarticletitle{Pervasive label errors in test sets destabilize machine learning benchmarks}.
\newblock \bibinfo{journal}{\emph{arXiv preprint arXiv:2103.14749}} (\bibinfo{year}{2021}).
\newblock


\bibitem[OpenAI(2025)]%
        {openai-gpt4-1}
\bibfield{author}{\bibinfo{person}{OpenAI}.} \bibinfo{year}{2025}\natexlab{}.
\newblock \bibinfo{booktitle}{\emph{Introducing GPT-4.1 in the API}}.
\newblock
\urldef\tempurl%
\url{https://openai.com/index/gpt-4-1}
\showURL{%
\tempurl}


\bibitem[Ouyang et~al\mbox{.}(2022)]%
        {ouyang2022training}
\bibfield{author}{\bibinfo{person}{Long Ouyang}, \bibinfo{person}{Jeffrey Wu}, \bibinfo{person}{Xu Jiang}, \bibinfo{person}{Diogo Almeida}, \bibinfo{person}{Carroll Wainwright}, \bibinfo{person}{Pamela Mishkin}, \bibinfo{person}{Chong Zhang}, \bibinfo{person}{Sandhini Agarwal}, \bibinfo{person}{Katarina Slama}, \bibinfo{person}{Alex Ray}, {et~al\mbox{.}}} \bibinfo{year}{2022}\natexlab{}.
\newblock \showarticletitle{Training language models to follow instructions with human feedback}.
\newblock \bibinfo{journal}{\emph{Advances in neural information processing systems}}  \bibinfo{volume}{35} (\bibinfo{year}{2022}), \bibinfo{pages}{27730--27744}.
\newblock


\bibitem[Pankajakshan et~al\mbox{.}(2024)]%
        {pankajakshan2024mapping}
\bibfield{author}{\bibinfo{person}{Rahul Pankajakshan}, \bibinfo{person}{Sumitra Biswal}, \bibinfo{person}{Yuvaraj Govindarajulu}, {and} \bibinfo{person}{Gilad Gressel}.} \bibinfo{year}{2024}\natexlab{}.
\newblock \showarticletitle{Mapping llm security landscapes: A comprehensive stakeholder risk assessment proposal}.
\newblock \bibinfo{journal}{\emph{arXiv preprint arXiv:2403.13309}} (\bibinfo{year}{2024}).
\newblock


\bibitem[Papernot et~al\mbox{.}(2016)]%
        {papernot2016limitations}
\bibfield{author}{\bibinfo{person}{Nicolas Papernot}, \bibinfo{person}{Patrick McDaniel}, \bibinfo{person}{Somesh Jha}, \bibinfo{person}{Matt Fredrikson}, \bibinfo{person}{Z~Berkay Celik}, {and} \bibinfo{person}{Ananthram Swami}.} \bibinfo{year}{2016}\natexlab{}.
\newblock \showarticletitle{The limitations of deep learning in adversarial settings}. In \bibinfo{booktitle}{\emph{2016 IEEE European symposium on security and privacy (EuroS\&P)}}. IEEE, \bibinfo{pages}{372--387}.
\newblock


\bibitem[Parraga et~al\mbox{.}(2025)]%
        {parraga2025fairness}
\bibfield{author}{\bibinfo{person}{Otavio Parraga}, \bibinfo{person}{Martin~D More}, \bibinfo{person}{Christian~M Oliveira}, \bibinfo{person}{Nathan~S Gavenski}, \bibinfo{person}{Lucas~S Kupssinsk{\"u}}, \bibinfo{person}{Adilson Medronha}, \bibinfo{person}{Luis~V Moura}, \bibinfo{person}{Gabriel~S Sim{\~o}es}, {and} \bibinfo{person}{Rodrigo~C Barros}.} \bibinfo{year}{2025}\natexlab{}.
\newblock \showarticletitle{Fairness in Deep Learning: A survey on vision and language research}.
\newblock \bibinfo{journal}{\emph{Comput. Surveys}} \bibinfo{volume}{57}, \bibinfo{number}{6} (\bibinfo{year}{2025}), \bibinfo{pages}{1--40}.
\newblock


\bibitem[Ramirez et~al\mbox{.}(2022)]%
        {ramirez2022poisoning}
\bibfield{author}{\bibinfo{person}{Miguel~A Ramirez}, \bibinfo{person}{Song-Kyoo Kim}, \bibinfo{person}{Hussam~Al Hamadi}, \bibinfo{person}{Ernesto Damiani}, \bibinfo{person}{Young-Ji Byon}, \bibinfo{person}{Tae-Yeon Kim}, \bibinfo{person}{Chung-Suk Cho}, {and} \bibinfo{person}{Chan~Yeob Yeun}.} \bibinfo{year}{2022}\natexlab{}.
\newblock \showarticletitle{Poisoning attacks and defenses on artificial intelligence: A survey}.
\newblock \bibinfo{journal}{\emph{arXiv preprint arXiv:2202.10276}} (\bibinfo{year}{2022}).
\newblock


\bibitem[Rigaki and Garcia(2023)]%
        {rigaki2023survey}
\bibfield{author}{\bibinfo{person}{Maria Rigaki} {and} \bibinfo{person}{Sebastian Garcia}.} \bibinfo{year}{2023}\natexlab{}.
\newblock \showarticletitle{A survey of privacy attacks in machine learning}.
\newblock \bibinfo{journal}{\emph{Comput. Surveys}} \bibinfo{volume}{56}, \bibinfo{number}{4} (\bibinfo{year}{2023}), \bibinfo{pages}{1--34}.
\newblock


\bibitem[Saha et~al\mbox{.}(2020)]%
        {saha2020hidden}
\bibfield{author}{\bibinfo{person}{Aniruddha Saha}, \bibinfo{person}{Akshayvarun Subramanya}, {and} \bibinfo{person}{Hamed Pirsiavash}.} \bibinfo{year}{2020}\natexlab{}.
\newblock \showarticletitle{Hidden trigger backdoor attacks}. In \bibinfo{booktitle}{\emph{Proceedings of the AAAI conference on artificial intelligence}}, Vol.~\bibinfo{volume}{34}. \bibinfo{pages}{11957--11965}.
\newblock


\bibitem[Schmitt and Flechais(2024)]%
        {schmitt2024digital}
\bibfield{author}{\bibinfo{person}{Marc Schmitt} {and} \bibinfo{person}{Ivan Flechais}.} \bibinfo{year}{2024}\natexlab{}.
\newblock \showarticletitle{Digital deception: Generative artificial intelligence in social engineering and phishing}.
\newblock \bibinfo{journal}{\emph{Artificial Intelligence Review}} \bibinfo{volume}{57}, \bibinfo{number}{12} (\bibinfo{year}{2024}), \bibinfo{pages}{1--23}.
\newblock


\bibitem[Schr{\"o}er et~al\mbox{.}(2025)]%
        {schroer2025sok}
\bibfield{author}{\bibinfo{person}{Saskia~Laura Schr{\"o}er}, \bibinfo{person}{Giovanni Apruzzese}, \bibinfo{person}{Soheil Human}, \bibinfo{person}{Pavel Laskov}, \bibinfo{person}{Hyrum~S Anderson}, \bibinfo{person}{Edward~WN Bernroider}, \bibinfo{person}{Aurore Fass}, \bibinfo{person}{Ben Nassi}, \bibinfo{person}{Vera Rimmer}, \bibinfo{person}{Fabio Roli}, {et~al\mbox{.}}} \bibinfo{year}{2025}\natexlab{}.
\newblock \showarticletitle{SoK: On the offensive potential of AI}. In \bibinfo{booktitle}{\emph{2025 IEEE Conference on Secure and Trustworthy Machine Learning (SaTML)}}. IEEE, \bibinfo{pages}{247--280}.
\newblock


\bibitem[Schwalbe and Finzel(2024)]%
        {schwalbe2024comprehensive}
\bibfield{author}{\bibinfo{person}{Gesina Schwalbe} {and} \bibinfo{person}{Bettina Finzel}.} \bibinfo{year}{2024}\natexlab{}.
\newblock \showarticletitle{A comprehensive taxonomy for explainable artificial intelligence: a systematic survey of surveys on methods and concepts}.
\newblock \bibinfo{journal}{\emph{Data Mining and Knowledge Discovery}} \bibinfo{volume}{38}, \bibinfo{number}{5} (\bibinfo{year}{2024}), \bibinfo{pages}{3043--3101}.
\newblock


\bibitem[Shankar et~al\mbox{.}(2017)]%
        {shankar2017no}
\bibfield{author}{\bibinfo{person}{Shreya Shankar}, \bibinfo{person}{Yoni Halpern}, \bibinfo{person}{Eric Breck}, \bibinfo{person}{James Atwood}, \bibinfo{person}{Jimbo Wilson}, {and} \bibinfo{person}{D Sculley}.} \bibinfo{year}{2017}\natexlab{}.
\newblock \showarticletitle{No classification without representation: Assessing geodiversity issues in open data sets for the developing world}.
\newblock \bibinfo{journal}{\emph{arXiv preprint arXiv:1711.08536}} (\bibinfo{year}{2017}).
\newblock


\bibitem[Shen et~al\mbox{.}(2022)]%
        {shen2022sok}
\bibfield{author}{\bibinfo{person}{Junjie Shen}, \bibinfo{person}{Ningfei Wang}, \bibinfo{person}{Ziwen Wan}, \bibinfo{person}{Yunpeng Luo}, \bibinfo{person}{Takami Sato}, \bibinfo{person}{Zhisheng Hu}, \bibinfo{person}{Xinyang Zhang}, \bibinfo{person}{Shengjian Guo}, \bibinfo{person}{Zhenyu Zhong}, \bibinfo{person}{Kang Li}, {et~al\mbox{.}}} \bibinfo{year}{2022}\natexlab{}.
\newblock \showarticletitle{Sok: On the semantic ai security in autonomous driving}.
\newblock \bibinfo{journal}{\emph{arXiv preprint arXiv:2203.05314}} (\bibinfo{year}{2022}).
\newblock


\bibitem[Shen et~al\mbox{.}(2024)]%
        {shen2024anything}
\bibfield{author}{\bibinfo{person}{Xinyue Shen}, \bibinfo{person}{Zeyuan Chen}, \bibinfo{person}{Michael Backes}, \bibinfo{person}{Yun Shen}, {and} \bibinfo{person}{Yang Zhang}.} \bibinfo{year}{2024}\natexlab{}.
\newblock \showarticletitle{" do anything now": Characterizing and evaluating in-the-wild jailbreak prompts on large language models}. In \bibinfo{booktitle}{\emph{Proceedings of the 2024 on ACM SIGSAC Conference on Computer and Communications Security}}. \bibinfo{pages}{1671--1685}.
\newblock


\bibitem[Shokri et~al\mbox{.}(2017)]%
        {shokri2017membership}
\bibfield{author}{\bibinfo{person}{Reza Shokri}, \bibinfo{person}{Marco Stronati}, \bibinfo{person}{Congzheng Song}, {and} \bibinfo{person}{Vitaly Shmatikov}.} \bibinfo{year}{2017}\natexlab{}.
\newblock \showarticletitle{Membership inference attacks against machine learning models}. In \bibinfo{booktitle}{\emph{2017 IEEE symposium on security and privacy (SP)}}. IEEE, \bibinfo{pages}{3--18}.
\newblock


\bibitem[Slattery et~al\mbox{.}(2024)]%
        {slattery2024ai}
\bibfield{author}{\bibinfo{person}{Peter Slattery}, \bibinfo{person}{Alexander~K Saeri}, \bibinfo{person}{Emily~AC Grundy}, \bibinfo{person}{Jess Graham}, \bibinfo{person}{Michael Noetel}, \bibinfo{person}{Risto Uuk}, \bibinfo{person}{James Dao}, \bibinfo{person}{Soroush Pour}, \bibinfo{person}{Stephen Casper}, {and} \bibinfo{person}{Neil Thompson}.} \bibinfo{year}{2024}\natexlab{}.
\newblock \showarticletitle{The ai risk repository: A comprehensive meta-review, database, and taxonomy of risks from artificial intelligence}.
\newblock \bibinfo{journal}{\emph{arXiv preprint arXiv:2408.12622}} (\bibinfo{year}{2024}).
\newblock


\bibitem[Slijep{\v{c}}evi{\'c} et~al\mbox{.}(2021)]%
        {slijepvcevic2021k}
\bibfield{author}{\bibinfo{person}{Djordje Slijep{\v{c}}evi{\'c}}, \bibinfo{person}{Maximilian Henzl}, \bibinfo{person}{Lukas~Daniel Klausner}, \bibinfo{person}{Tobias Dam}, \bibinfo{person}{Peter Kieseberg}, {and} \bibinfo{person}{Matthias Zeppelzauer}.} \bibinfo{year}{2021}\natexlab{}.
\newblock \showarticletitle{k-anonymity in practice: How generalisation and suppression affect machine learning classifiers}.
\newblock \bibinfo{journal}{\emph{Computers \& Security}}  \bibinfo{volume}{111} (\bibinfo{year}{2021}), \bibinfo{pages}{102488}.
\newblock


\bibitem[Szegedy et~al\mbox{.}(2013)]%
        {szegedy2013intriguing}
\bibfield{author}{\bibinfo{person}{Christian Szegedy}, \bibinfo{person}{Wojciech Zaremba}, \bibinfo{person}{Ilya Sutskever}, \bibinfo{person}{Joan Bruna}, \bibinfo{person}{Dumitru Erhan}, \bibinfo{person}{Ian Goodfellow}, {and} \bibinfo{person}{Rob Fergus}.} \bibinfo{year}{2013}\natexlab{}.
\newblock \showarticletitle{Intriguing properties of neural networks}.
\newblock \bibinfo{journal}{\emph{arXiv preprint arXiv:1312.6199}} (\bibinfo{year}{2013}).
\newblock


\bibitem[Tramer(2022)]%
        {tramer2022detecting}
\bibfield{author}{\bibinfo{person}{Florian Tramer}.} \bibinfo{year}{2022}\natexlab{}.
\newblock \showarticletitle{Detecting adversarial examples is (nearly) as hard as classifying them}. In \bibinfo{booktitle}{\emph{International conference on machine learning}}. PMLR, \bibinfo{pages}{21692--21702}.
\newblock


\bibitem[Tram{\`e}r et~al\mbox{.}(2016)]%
        {tramer2016stealing}
\bibfield{author}{\bibinfo{person}{Florian Tram{\`e}r}, \bibinfo{person}{Fan Zhang}, \bibinfo{person}{Ari Juels}, \bibinfo{person}{Michael~K Reiter}, {and} \bibinfo{person}{Thomas Ristenpart}.} \bibinfo{year}{2016}\natexlab{}.
\newblock \showarticletitle{Stealing machine learning models via prediction $\{$APIs$\}$}. In \bibinfo{booktitle}{\emph{25th USENIX security symposium (USENIX Security 16)}}. \bibinfo{pages}{601--618}.
\newblock


\bibitem[Valentin et~al\mbox{.}(2024)]%
        {valentin2024cost}
\bibfield{author}{\bibinfo{person}{Simon Valentin}, \bibinfo{person}{Jinmiao Fu}, \bibinfo{person}{Gianluca Detommaso}, \bibinfo{person}{Shaoyuan Xu}, \bibinfo{person}{Giovanni Zappella}, {and} \bibinfo{person}{Bryan Wang}.} \bibinfo{year}{2024}\natexlab{}.
\newblock \showarticletitle{Cost-effective hallucination detection for llms}.
\newblock \bibinfo{journal}{\emph{arXiv preprint arXiv:2407.21424}} (\bibinfo{year}{2024}).
\newblock


\bibitem[Wang et~al\mbox{.}(2020b)]%
        {wang2020federated}
\bibfield{author}{\bibinfo{person}{Hongyi Wang}, \bibinfo{person}{Mikhail Yurochkin}, \bibinfo{person}{Yuekai Sun}, \bibinfo{person}{Dimitris Papailiopoulos}, {and} \bibinfo{person}{Yasaman Khazaeni}.} \bibinfo{year}{2020}\natexlab{b}.
\newblock \showarticletitle{Federated learning with matched averaging}.
\newblock \bibinfo{journal}{\emph{arXiv preprint arXiv:2002.06440}} (\bibinfo{year}{2020}).
\newblock


\bibitem[Wang et~al\mbox{.}(2020a)]%
        {wang2020cnn}
\bibfield{author}{\bibinfo{person}{Sheng-Yu Wang}, \bibinfo{person}{Oliver Wang}, \bibinfo{person}{Richard Zhang}, \bibinfo{person}{Andrew Owens}, {and} \bibinfo{person}{Alexei~A Efros}.} \bibinfo{year}{2020}\natexlab{a}.
\newblock \showarticletitle{CNN-generated images are surprisingly easy to spot... for now}. In \bibinfo{booktitle}{\emph{Proceedings of the IEEE/CVF conference on computer vision and pattern recognition}}. \bibinfo{pages}{8695--8704}.
\newblock


\bibitem[Warnecke et~al\mbox{.}(2021)]%
        {warnecke2021machine}
\bibfield{author}{\bibinfo{person}{Alexander Warnecke}, \bibinfo{person}{Lukas Pirch}, \bibinfo{person}{Christian Wressnegger}, {and} \bibinfo{person}{Konrad Rieck}.} \bibinfo{year}{2021}\natexlab{}.
\newblock \showarticletitle{Machine unlearning of features and labels}.
\newblock \bibinfo{journal}{\emph{arXiv preprint arXiv:2108.11577}} (\bibinfo{year}{2021}).
\newblock


\bibitem[Weidinger et~al\mbox{.}(2022)]%
        {weidinger2022taxonomy}
\bibfield{author}{\bibinfo{person}{Laura Weidinger}, \bibinfo{person}{Jonathan Uesato}, \bibinfo{person}{Maribeth Rauh}, \bibinfo{person}{Conor Griffin}, \bibinfo{person}{Po-Sen Huang}, \bibinfo{person}{John Mellor}, \bibinfo{person}{Amelia Glaese}, \bibinfo{person}{Myra Cheng}, \bibinfo{person}{Borja Balle}, \bibinfo{person}{Atoosa Kasirzadeh}, {et~al\mbox{.}}} \bibinfo{year}{2022}\natexlab{}.
\newblock \showarticletitle{Taxonomy of risks posed by language models}. In \bibinfo{booktitle}{\emph{Proceedings of the 2022 ACM conference on fairness, accountability, and transparency}}. \bibinfo{pages}{214--229}.
\newblock


\bibitem[Wingarz et~al\mbox{.}(2024)]%
        {wingarz2024sok}
\bibfield{author}{\bibinfo{person}{Tatjana Wingarz}, \bibinfo{person}{Anne Lauscher}, \bibinfo{person}{Janick Edinger}, \bibinfo{person}{Dominik Kaaser}, \bibinfo{person}{Stefan Schulte}, {and} \bibinfo{person}{Mathias Fischer}.} \bibinfo{year}{2024}\natexlab{}.
\newblock \showarticletitle{SoK: Towards Security and Safety of Edge AI}.
\newblock \bibinfo{journal}{\emph{arXiv preprint arXiv:2410.05349}} (\bibinfo{year}{2024}).
\newblock


\bibitem[Xu et~al\mbox{.}(2021)]%
        {xu2021detecting}
\bibfield{author}{\bibinfo{person}{Xiaojun Xu}, \bibinfo{person}{Qi Wang}, \bibinfo{person}{Huichen Li}, \bibinfo{person}{Nikita Borisov}, \bibinfo{person}{Carl~A Gunter}, {and} \bibinfo{person}{Bo Li}.} \bibinfo{year}{2021}\natexlab{}.
\newblock \showarticletitle{Detecting ai trojans using meta neural analysis}. In \bibinfo{booktitle}{\emph{2021 IEEE Symposium on Security and Privacy (SP)}}. IEEE, \bibinfo{pages}{103--120}.
\newblock


\bibitem[Ye et~al\mbox{.}(2024)]%
        {ye2024benchmarking}
\bibfield{author}{\bibinfo{person}{Fanghua Ye}, \bibinfo{person}{Mingming Yang}, \bibinfo{person}{Jianhui Pang}, \bibinfo{person}{Longyue Wang}, \bibinfo{person}{Derek Wong}, \bibinfo{person}{Emine Yilmaz}, \bibinfo{person}{Shuming Shi}, {and} \bibinfo{person}{Zhaopeng Tu}.} \bibinfo{year}{2024}\natexlab{}.
\newblock \showarticletitle{Benchmarking llms via uncertainty quantification}.
\newblock \bibinfo{journal}{\emph{Advances in Neural Information Processing Systems}}  \bibinfo{volume}{37} (\bibinfo{year}{2024}), \bibinfo{pages}{15356--15385}.
\newblock


\bibitem[Zhang et~al\mbox{.}(2024b)]%
        {zhang2024safeguard}
\bibfield{author}{\bibinfo{person}{Qingzhao Zhang}, \bibinfo{person}{Ziyang Xiong}, {and} \bibinfo{person}{Z~Morley Mao}.} \bibinfo{year}{2024}\natexlab{b}.
\newblock \showarticletitle{Safeguard is a Double-edged Sword: Denial-of-service Attack on Large Language Models}.
\newblock \bibinfo{journal}{\emph{arXiv preprint arXiv:2410.02916}} (\bibinfo{year}{2024}).
\newblock


\bibitem[Zhang et~al\mbox{.}(2024a)]%
        {zhang2024backdoor}
\bibfield{author}{\bibinfo{person}{Shaobo Zhang}, \bibinfo{person}{Yimeng Pan}, \bibinfo{person}{Qin Liu}, \bibinfo{person}{Zheng Yan}, \bibinfo{person}{Kim-Kwang~Raymond Choo}, {and} \bibinfo{person}{Guojun Wang}.} \bibinfo{year}{2024}\natexlab{a}.
\newblock \showarticletitle{Backdoor attacks and defenses targeting multi-domain ai models: A comprehensive review}.
\newblock \bibinfo{journal}{\emph{Comput. Surveys}} \bibinfo{volume}{57}, \bibinfo{number}{4} (\bibinfo{year}{2024}), \bibinfo{pages}{1--35}.
\newblock


\bibitem[Zhang(2024)]%
        {zhang2024ai}
\bibfield{author}{\bibinfo{person}{Wenbin Zhang}.} \bibinfo{year}{2024}\natexlab{}.
\newblock \showarticletitle{AI fairness in practice: Paradigm, challenges, and prospects}.
\newblock \bibinfo{journal}{\emph{Ai Magazine}} \bibinfo{volume}{45}, \bibinfo{number}{3} (\bibinfo{year}{2024}), \bibinfo{pages}{386--395}.
\newblock


\bibitem[Zhang et~al\mbox{.}(2020)]%
        {zhang2020secret}
\bibfield{author}{\bibinfo{person}{Yuheng Zhang}, \bibinfo{person}{Ruoxi Jia}, \bibinfo{person}{Hengzhi Pei}, \bibinfo{person}{Wenxiao Wang}, \bibinfo{person}{Bo Li}, {and} \bibinfo{person}{Dawn Song}.} \bibinfo{year}{2020}\natexlab{}.
\newblock \showarticletitle{The secret revealer: Generative model-inversion attacks against deep neural networks}. In \bibinfo{booktitle}{\emph{Proceedings of the IEEE/CVF conference on computer vision and pattern recognition}}. \bibinfo{pages}{253--261}.
\newblock


\end{thebibliography}

\appendix

\section{Definition of Elements in Two Aspects} \label{apdx:def_aspects}
Table~\ref{tab:info_def} shows definitions of elements in the ISA.
\begin{table}[t]
    \centering
    \caption{Elements in the ISA}
    \scalebox{0.7}{
    \begin{tabular}{|c|p{0.5\textwidth}|}
    \hline
        \textbf{Elements that } & \multirow{2}{*}{\textbf{Definition}} \\
        \textbf{AI should satisfy} & \\ \hline
        Confidentiality & AI data and models are not accessed by unauthorized individuals. \\\hline
        \multirow{2}{*}{Integrity} & The AI models and algorithms have not been tampered with, and the AI outputs are as expected. \\\hline
        Availability & AI can provide the necessary features and services when needed. \\\hline
        Explainability & AI can explain the basis and process of its output. \\\hline
        \multirow{2}{*}{Output Fairness} & AI does not produce biased outputs towards specific individuals or groups. \\\hline
        \multirow{2}{*}{Safety} & AI is equipped with safety mechanisms to prevent harm to human life, body, property, or mind.\\\hline
        Accuracy & AI meets a certain level of accuracy for achieving objectives. \\\hline
        \multirow{2}{*}{Controllability} & AI is controlled by administrators and does not run amok or affect other environments. \\\hline
        Trustworthiness & Output from AI is reliable. \\\hline
    \end{tabular}
    }
    \label{tab:info_def}
\end{table}
Table~\ref{tab:ext_def} shows definitions of elements in the EIA.
\begin{table}[t]
    \centering
    \caption{Elements in the EIA}
    
    \scalebox{0.7}{
    \begin{tabular}{|c|p{0.45\textwidth}|}
    \hline
        \textbf{Elements that impact} & \multirow{2}{*}{\textbf{Definition}} \\
        \textbf{individuals and society} &  \\ \hline
        Cyber attack & AI is not used for cyber attacks. \\\hline
        Military use & AI is not used for military purposes.\\\hline
        \multirow{2}{*}{Privacy} & AI does not infringe on privacy and complies with privacy laws and customs.\\\hline
        \multirow{2}{*}{Disinformation} & AI is not used to intentionally create disinformation, or it can identify such disinformation.\\\hline
        \multirow{2}{*}{Misinformation} & AI does not output misinformation, or it can identify such misinformation.\\\hline
        Usability & AI meets a certain level of usability to achieve objectives.\\\hline
        Consumer fairness & No harm is caused by unfair biased output from AI. \\\hline
        Plagiarism & AI is not used for plagiarism. \\\hline
        \multirow{2}{*}{Copyright and authorship} & AI complies with laws and customs related to copyright and authorship. \\\hline
        \multirow{2}{*}{Transparency} & It is clearly stated that the system uses AI, including information on its limitations and risks associated with its use. \\\hline
        \multirow{2}{*}{Reputation} & The AI system provider is evaluated to a certain standard and is trusted. \\ \hline
        \multirow{2}{*}{\makecell{Compliance with \\ laws and regulations}} & AI is used for lawful purposes and produces output or actions that comply with the law. \\ \hline
        Human-centric principle & AI is appropriately used for the benefit of humans. \\\hline
        Ethics & AI behaves in a manner consistent with societal norms. \\\hline
        Economy & The use of AI has a positive impact on economy. \\\hline
        Physical impact & The use of AI does not cause physical harm to people. \\\hline
        Psychological impact & The use of AI does not cause psychological harm to people. \\\hline
        Financial impact & The use of AI does not cause financial harm to people. \\ \hline
        \multirow{2}{*}{Medical care} & The use of AI contributes to the development of advanced and safe medical care. \\ \hline
        \multirow{2}{*}{Critical infrastructure} & The use of AI contributes to the safe operation of critical infrastructure. \\ 
        \hline
    \end{tabular}
    }
    \label{tab:ext_def}
\end{table}

\section{Definition of security targets in our work} \label{apdx:def_target}
Table~\ref{tab:security_target} shows definition of security targets in our work.
\begin{table}[t]
    \centering
    \caption{Security targets in AI security map}
    \scalebox{0.7}{
    \begin{tabular}{|c|p{0.5\textwidth}|}
    \hline
    \textbf{Security target}& \textbf{Definition} \\ \hline
        Consumer & An individual or organization that utilizes AI or AI systems. \\ \hline
        Non-consumer & An individual or organization that is not classified as a consumer.\\ \hline
        Society	& A group composed of multiple people and organizations.\\ \hline
        \multirow{2}{*}{AI system provider} & An individual or organization that provides an information system using AI (AI system). \\ \hline
    \end{tabular}
    }
    \label{tab:security_target}
\end{table}


\section{Prompt example} \label{prompt_example}
The following example is a prompt used for the trial in Section~\ref{sec:recommend}.
\\
\\
\\
\\
\\
\begin{boxprompt}
\footnotesize
Analyze the following news article and extract the negative impacts of AI technology on individuals, organizations, or society that are explicitly stated in the article based on the specified list of negative impacts. 
If a negative impact is not yet apparent and there is only a possibility the negative impact may occur in the future, it should be excluded. 
The extracted negative impacts should be output in JSON format.

\# List of Negative Impacts: \\
- Decrease in AI prediction accuracy \\
- Personal information leakage \\
- Using AI for cyber attacks \\
.... \\
\# News Article \\
{[}input news here{]}\\

\# Output Format:

\{ \\
  "Negative impacts": {[} \\
    \{
      "impact": "[Negative impact]", \\
      "description": "[Explanation or relevant part of the news article]" 
    \}, \\
    // Add other negative impacts as needed \\
  {]}
\}
\end{boxprompt}

\end{document}